\documentclass[prb,showpacs,preprint]{revtex4}
\usepackage{amsmath}
\usepackage{graphicx}
\begin{document}
\title{Electron-vibron coupling effects on electron transport via a single-molecule magnet}
\author{Alexander McCaskey$^1$, Yoh Yamamoto$^1$, Michael Warnock$^1$, Enrique Burzur\'i$^2$,
Herre S. J. van der Zant$^2$, and Kyungwha Park$^1$} \email{kyungwha@vt.edu}
\affiliation{$^1$Department of Physics, Virginia Tech, Blacksburg, Virginia 24061, USA \\
$^2$Kavli Institute of Nanoscience, Delft University of Technology, P.O. Box 5046,
2600 GA Delft, The Netherlands}
\begin{abstract}
We investigate how the electron-vibron coupling influences electron transport via an anisotropic magnetic molecule, such as a
single-molecule magnet (SMM) Fe$_4$, by using a model Hamiltonian with parameter values obtained from density-functional theory (DFT).
Magnetic anisotropy parameters, vibrational energies, and electron-vibron coupling strengths of the Fe$_4$ are computed using DFT.
A giant spin model is applied to the Fe$_4$ with only two charge states, specifically a neutral state with the total spin $S=5$
and a singly charged state with $S=9/2$, which is consistent with our DFT result and experiments on Fe$_4$ single-molecule transistors.
In sequential electron tunneling, we find that the magnetic anisotropy gives rise to new features in conductance peaks arising from 
vibrational excitations. In particular, the peak height shows a strong, unusual dependence on the {\it direction} as well as magnitude of
applied $B$ field. The magnetic anisotropy also introduces vibrational satellite peaks whose position and height are modified
with the direction and magnitude of applied $B$ field. Furthermore, when multiple vibrational modes with considerable
electron-vibron coupling have energies close to one another, a low-bias current is suppressed, independently of gate voltage
and applied $B$ field, although that is not the case for a single mode with the similar electron-vibron coupling. In the former case,
the conductance peaks reveal a stronger $B$-field dependence than in the latter case. The new features appear because
the magnetic anisotropy barrier is of the same order of magnitude as the energies of vibrational modes with
significant electron-vibron coupling. Our findings clearly show the interesting interplay between magnetic anisotropy and
electron-vibron coupling in electron transport via the Fe$_4$. The similar behavior can be observed in transport via other anisotropic
magnetic molecules.
\end{abstract}
\date{\today}

\maketitle

\section{Introduction}



Recent experimental advances allow individual molecules to be placed between electrodes, and their electron transport properties
to be measured in single-molecule junctions or transistors. One interesting family of molecules among them are anisotropic magnetic
molecules referred to as single-molecule magnets (SMMs). A SMM comprises a few transition metal ions surrounded by several tens to hundreds
of atoms, and has a large spin and a large magnetic anisotropy barrier \cite{FRIE96,THOM96,CHUD98}. Crystals of SMMs have drawn
attention due to unique quantum properties such as quantum tunneling of magnetization \cite{FRIE96,THOM96} and quantum interference
or Berry-phase oscillations induced by the magnetic anisotropy \cite{WERN99,GARG93,BURZ13}. There have been studies of the interplay 
between the quantum properties and the electron transport of individual SMMs at the single-molecule level
\cite{HEER06,LEUE06,ROME06,TIMM06,JO06,SALV09,ZYAZ10,CHUD11,TIMM12,GANZ13,MISI13,WU13,ROME14}.

Molecules trapped in single-molecule devices vibrate with discrete frequencies characteristic to the molecules, and the molecular
vibrations can couple to electronic charge and/or spin degrees of freedom. When this coupling is significant, electrons may tunnel via
the vibrational excitations unique to the molecules, and the coupling can be tailored by external means. Electron tunneling through vibrational
excitations have been observed in single-molecule devices based on carbon nanotubes \cite{GANZ13,LERO04,SAPM06,LETU09,BENY14} and small
molecules \cite{STIP98,PARK00,YU04,LEON08} including SMMs such as Fe$_4$ \cite{BURZ14}. Interestingly, in some cases, a pronounced
suppression of a low-bias current was found, attributed to a strong coupling between electronic charge and vibrations of
nanosystems \cite{SAPM06,LETU09,BURZ14,KOCH05,KOCH06}. It was also shown that the coupling strength could be modified at the nanometer
scale in carbon nanotube mechanical resonators \cite{BENY14}. For a SMM TbPc$_2$ grafted onto a carbon nanotube, a coupling between
the molecular spin and vibrations of the nanotube was observed in conductance maps of the nanotube \cite{GANZ13}.

So far, theories of the electron-phonon or electron-vibron coupling effects have been developed only for isotropic molecules
\cite{KOCH06,MITR04,GALP07,SECK11,HART13,CORN07,FRED07,PAUL05,MCCA03,SELD08} in single-molecule junctions
or transistors. For example, for molecules weakly coupled to electrodes, a model Hamiltonian approach is commonly used to investigate the
coupling effects, while for molecules strongly coupled to electrodes, a first-principles based method such as density-functional theory
(DFT) combined with non-equilibrium Green's function method, is applied \cite{FRED07}. Recently, the coupling effects have been studied for
isotropic molecules weakly coupled to electrodes, by using both DFT and the model Hamiltonian approach \cite{SELD08}. For
anisotropic magnetic molecules weakly coupled to electrodes, a combination of DFT and a model Hamiltonian would be proper to
examine the coupling effects. The interplay between magnetic anisotropy and vibron-assisted tunneling can provide interesting features
concerning vibrational conductance peaks.

\begin{figure}[h]
\includegraphics[width=0.9\textwidth]{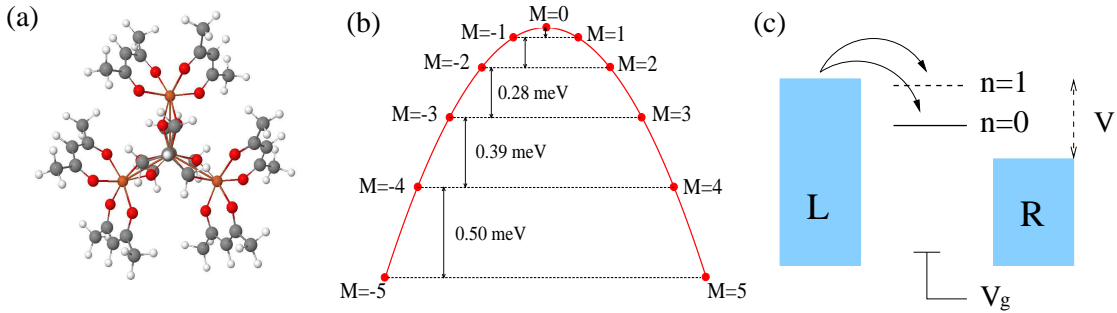}
\caption{(Color online) (a) Top view of the Fe$_4$ molecule with $C_2$ symmetry axis along the vertical axis, where
Fe (orange), O (red), C (gray), H (white). Simplified from Ref.\cite{ACCO06}. (b) Magnetic energy levels of the Fe$_4$ with $S=5$
where the zero-field splitting is 0.50 meV. (c) Schematic view of sequential tunneling from the left electrode to a molecular level
($n=0$: vibrational ground state, $n=1$: vibrational first-excited state), where $V$ is a bias voltage and $V_g$ is a gate voltage.
The magnetic levels in each vibrational state are not shown. The chemical potential of the left and right electrodes are 
$+eV/2$ and $-eV/2$, respectively.}
\label{fig:geo}
\end{figure}

The SMM Fe$_4$ has been shown to form stable single-molecule transistors without linker groups \cite{ZYAZ10,BURZ12,BURZ14}.
The Fe$_4$ consists of four Fe$^{3+}$ ions (each ion with spin $S_i=5/2$), among which the center Fe$^{3+}$ ion is weakly antiferromagnetically
coupled to the outer Fe$^{3+}$ ions via O anions, as shown in Fig.~\ref{fig:geo}(a). The neutral Fe$_4$ has the total ground-state spin $S=5$
with a magnetic anisotropy barrier of 16.2~K [Fig.~\ref{fig:geo}(b)] \cite{ACCO06,BURZ12,BURZ14}, while its doubly degenerate excited spin
multiplets $S=4$ are located at 4.8~meV above the ground-state spin multiplet $S=5$ \cite{ACCO06}. The negatively singly charged Fe$_4$ has the
total spin $S=9/2$ well separated from the excited spin multiplet $S=11/2$. The previous DFT calculations suggest that the Fe$_4$ has
only three vibrational modes with the electron-vibron coupling greater than unity \cite{BURZ14}.

Here we present three electron-vibron coupling effects on electron transport via the SMM Fe$_4$ at low temperatures, in a sequential electron
tunneling limit [Fig.~\ref{fig:geo}(c)], by using the model Hamiltonian with the DFT-calculated magnetic anisotropy parameters, vibrational 
energies, and electron-vibron coupling strengths. Firstly, the height of vibrational conductance peaks shows a strong, unusual dependence on the
{\it direction} and magnitude of applied $B$ field. This $B$-field dependence is attributed to the magnetic anisotropy barrier that is of the
same order of magnitude as the energies of the vibrational modes with significant electron-vibron coupling. Without the magnetic anisotropy,
the conductance peaks would be insensitive to the $B$-field direction. Secondly, satellite conductance peaks of magnetic origin exhibit
a unique $B$-field evolution depending on the direction of $B$ field. At low $B$ fields, the low-bias satellite peak arises from the magnetic
levels in the vibrational ground state only, while at high $B$ fields, the levels in the vibrational excited states contribute to the
satellite peak as much as that those in the vibrational ground state, because the separation between the levels becomes comparable
to the vibrational excitations. Thirdly, when multiple modes with significant electron-vibron coupling ($1 < \lambda < 2$) have energies close to
one another, the low-bias conductance peak and the $B$-field dependence of the conductance peaks reveal qualitatively different features from
the case of a single mode with the similar electron-vibron coupling. The similar trend to our findings may be observed for any anisotropic
magnetic molecules as long as magnetic anisotropy is comparable to vibrational energies. This work can be viewed as a starting point
for an understanding of magnetic anisotropy effects on electron tunneling via vibrational excitations, by using the combined method.

The outline of this work is as follows. We present the DFT method in Sec.II, and show our DFT results on electronic structure and magnetic and
vibrational properties of the Fe$_4$ in Sec.III. We introduce the model Hamiltonian and a formalism for solving the master equation in Sec.IV, 
and discuss calculated transport properties of the Fe$_4$ as a function of gate voltage, temperature, and applied $B$ field in Sec.V.
Finally, we make a conclusion in Sec.VI.

\section{DFT calculation Method}

We perform electronic structure calculations of an isolated Fe$_4$ molecule using the DFT code, {\tt NRLMOL} \cite{NRLMOL}, considering
all electrons with Gaussian basis sets within the generalized-gradient approximation (GGA) \cite{PERD96} for the exchange-correlation
functional. To reduce the computational cost, the Fe$_4$ molecule \cite{ACCO06} is simplified by replacing the terminating CH$_3$ groups
by H atoms, and by substituting the phenyl rings (above and below the plane where the Fe ions are located) with H atoms. Figure 1(a) shows
the simplified Fe$_4$ molecule with $C_2$ symmetry. Without such simplification, vibrational modes would not be obtained within a reasonable
compute time. It is confirmed that this simplification does not affect much the electronic and magnetic properties of the Fe$_4$ molecule (Sec.III.A).
The phenyl rings are known to have high-frequency vibrational modes (about 600-1000 cm$^{-1}$) \cite{LIU90}, while the electron-vibron coupling
is significant for low-frequency vibrational modes. Therefore, the replacement of the phenyl rings by H would not affect our calculation of
electron-vibron coupling strengths for low-frequency vibrational modes. The total magnetic moments of the neutral and charged Fe$_4$ molecules
are initially set to 10~$\mu_B$ and 9~$\mu_B$,
respectively, and they remain the same after geometry relaxation. The geometries of the neutral and charged Fe$_4$ molecules are relaxed with
$C_2$ symmetry, until the maximum force is less than 0.009 eV/\AA, or 0.00018 Ha/$a_B$, where $a_B$ is Bohr radius. For the relaxed geometry of
the neutral Fe$_4$, we calculate vibrational or normal modes within the harmonic oscillator approximation, using the frozen phonon method
\cite{NRLMOL}. We also calculate the magnetic anisotropy parameters for the neutral Fe$_4$ molecule by considering spin-orbit coupling
perturbatively to the converged Kohn-Sham orbitals and orbital energies obtained from DFT, as implemented in {\tt NRLMOL} \cite{NRLMOL,PEDE99}.

\section{DFT results: electronic, magnetic and vibrational properties}

\subsection{Electronic and magnetic properties}

\begin{figure}[h]
\includegraphics[width=0.6\textwidth]{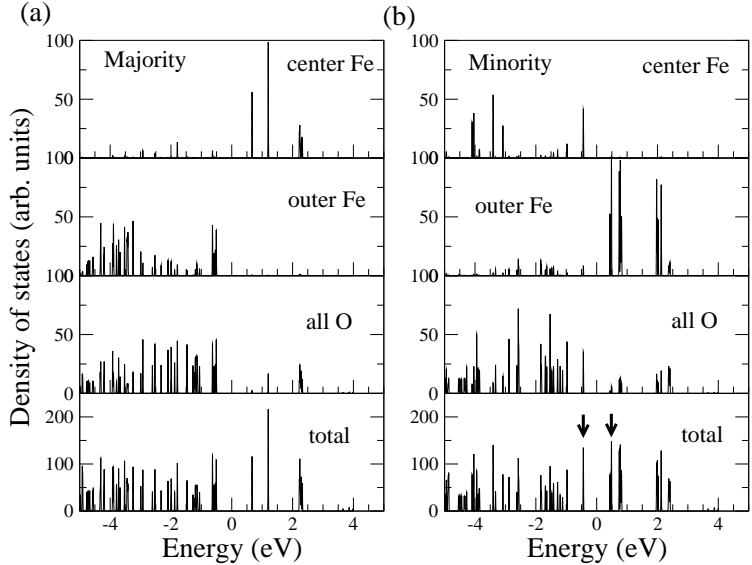}
\caption{(a) Majority- and (b) minority-spin total and projected density of states onto the center Fe and outer Fe sites and onto all
O atoms of the Fe$_4$ molecule shown in Fig.~\ref{fig:geo}(a). The midpoint between the HOMO and the LUMO levels is set to zero. 
The arrows in the bottom panel of (b)
indicate the HOMO and LUMO levels. Obtained from the neutral Fe$_4$.}
\label{fig:dos}
\end{figure}

Our DFT calculations show that the neutral Fe$_4$ molecule with $S=5$ has an energy gap of 0.87 eV between the lowest unoccupied molecular
orbital (LUMO) and the highest occupied molecular orbital (HOMO) levels. The HOMO level is doubly degenerate, while the doubly degenerate
LUMO+1 level is separated from the LUMO level by 0.05~eV. 
The LUMO arises from the outer Fe ions with the minority spin
(spin down) at the vertices of the triangle [Fig.~\ref{fig:geo}(a)], while the HOMO from the center Fe ion with the minority spin, as shown
in Fig.~\ref{fig:dos}. The O orbital levels are found at the same energies as the Fe orbital levels. The contributions of the C and H atoms
to the HOMO and LUMO are negligible. The majority-spin HOMO is 0.08~eV below the minority-spin HOMO, and the majority-spin LUMO is 0.23~eV
above the minority-spin LUMO. The calculated electronic structure suggests that when an extra electron is added to the Fe$_4$ molecule, the
electron is likely to go to the minority-spin outer Fe sites. Thus, the total spin of the charged Fe$_4$ is expected to be $S=9/2$, which is
consistent with our DFT calculation and experimental data \cite{BURZ12}. Furthermore, we calculate the uniaxial ($D$) and transverse magnetic anisotropy ($E$) parameters for the neutral Fe$_4$, finding
that $D$=0.056 meV and $E$=0.002~meV, respectively. These values are in good agreement with the experimental values, $D=0.056$ and $E=0.003$~meV \cite{BURZ12} and the previous DFT-calculated result \cite{NOSS13}. The calculated magnetic anisotropy barrier for the neutral Fe$_4$ is 16.2~K ($\sim$1.4~meV) [Fig.~\ref{fig:geo}(b)], in good agreement with experiment \cite{ACCO06,BURZ12}. The calculated zero-field splitting is 0.5~meV,
which is an energy difference between the two lowest doublets in the absence of external $B$ field.

The electronic structure study of the charged Fe$_4$ molecule, however, provides a HOMO-LUMO gap of 0.06 eV, which agrees with the previous DFT
result \cite{NOSS13}. This small gap is partially due to the degenerate LUMO levels and partially attributed to delocalization of the extra
electron over the Fe$_4$ (or difficulty in localization of the extra electron). The latter arises from an inherent limitation of DFT
caused by the absence of self-interaction corrections \cite{PEDE85}. The magnetic anisotropy parameters are highly sensitive to the HOMO-LUMO
gap and the location of the extra electron in the Fe$_4$. Therefore, in our transport calculations (Sec.V), for the charged Fe$_4$ molecule, we
use the DFT-calculated relaxed geometry but not the DFT-calculated magnetic anisotropy parameter values.

\subsection{Vibrational spectra and electron-vibron coupling}

\begin{figure}[h]
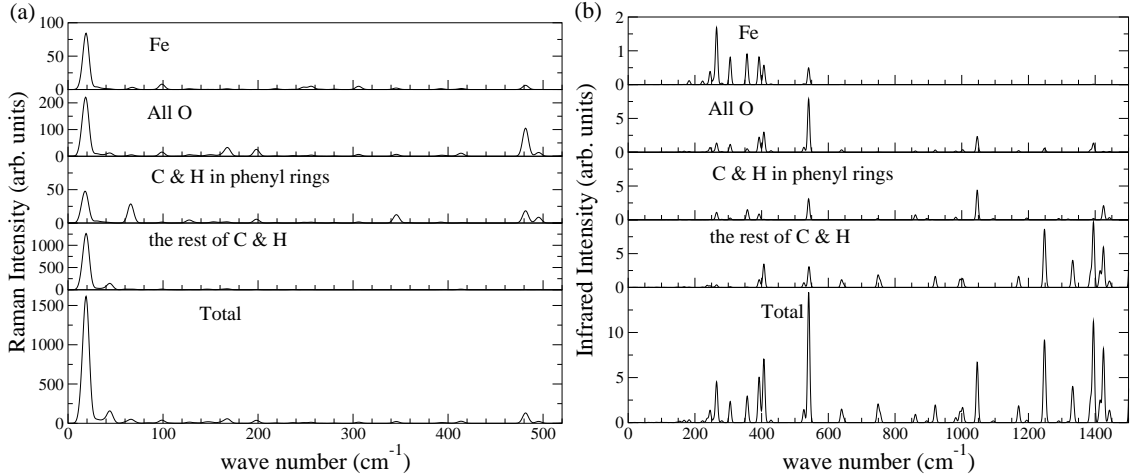

\includegraphics[width=0.45\textwidth]{Raman_spectra_v02.eps}
\includegraphics[width=0.45\textwidth]{IR_spectra_v02.eps}
\caption{Calculated (a) Raman and (b) infrared vibrational spectra of the neutral Fe$_4$ molecule with projections onto all Fe, all O,
all C and H atoms replacing the phenyl rings, and the peripheral C and H atoms.
The scales of the horizontal axes in (a) differ from those in (b).}
\label{fig:raman}
\end{figure}

We obtain total and projected Raman and infrared spectra by applying the scheme in Ref.~\cite{PORE96} to the DFT-calculated vibrational modes
of the neutral Fe$_4$ (Fig.~\ref{fig:raman}). There are 16 non-zero frequency normal modes below 50~cm$^{-1}$ (or 6.2 meV), among
which the lowest-energy mode has a frequency of 14.7~cm$^{-1}$. These low-frequency modes are all Raman active
[Fig.~\ref{fig:raman}(a)], and they involve with vibrations of Fe atoms and O and C atoms in the peripheral area. We compare our calculated
Raman spectra with experimental data in Ref.~\cite{BOGA09}. The experimental Raman spectrum is for a crystal of Fe$_4$ molecules with slightly
different ligands and only for high-frequency modes ($ > 200$~cm$^{-1}$). The experimental Raman peaks appear at 257, 378, 401, 413, 511, 539,
590~cm$^{-1}$, and they are all involved with Fe-O-Fe vibrations or stretch. The corresponding DFT Raman peaks are found at 255, 345, 393, 414,
482, 542~cm$^{-1}$, except for 590~cm$^{-1}$. Note that these peaks have much lower intensities than the 16 lower-frequency modes, so that some
of them are not visible in the scales of Fig.~\ref{fig:raman}(a). Infrared-active modes [Fig.~\ref{fig:raman}(b)] have much higher frequencies
than the Raman active modes.

\begin{figure}[h]
\includegraphics[width=0.95\textwidth]{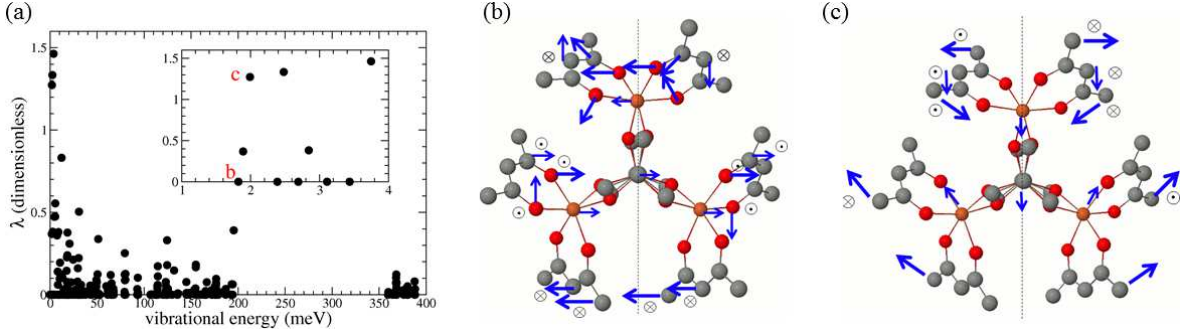}
\caption{(Color online) (a) Calculated electron-vibron coupling strength vs vibrational energy with an inset of several low-frequency
vibrational modes. (b) Vibrational mode ``b'' and (c) vibrational mode ``c'' marked in (a), where the arrows represent in-plane displacements
and $\bigodot$ and $\bigotimes$ are the positive and negative out-of-plane displacements, respectively. In (b) and (c), the vertical
dashed lines are the C$_2$ symmetry axes. (a) and (c) are adapted from Ref.~\cite{BURZ14}.}
\label{fig:ev}
\end{figure}

For each vibrational mode, the dimensionless electron-vibron coupling strength is given by \cite{MCCA03,SELD08,BURZ14}
\begin{eqnarray}
\lambda &=& \sqrt{\frac{\omega}{2 \hbar}} \mathbf{\Omega}^T \mathbf{M} (\mathbf{R}_0 - \mathbf{R}_1),
\label{eq:ev}
\end{eqnarray}
where $\omega$ is the angular frequency of the mode, $\mathbf{M}$ is a diagonal square matrix of atomic masses, and $\mathbf{\Omega}^T$ is
a transpose of the mass-weighted normal-mode column eigenvector with $\mathbf{\Omega}^T \mathbf{M} \mathbf{\Omega}=1$. Here $\mathbf{R}_0$ and
$\mathbf{R}_1$ are column vectors representing the coordinates of the neutral and charged Fe$_4$ relaxed geometries, respectively. The relaxed
geometries are translated and rotated such that $|\mathbf{R}_0 - \mathbf{R}_1|$ is minimized. Figure~\ref{fig:ev}(a) shows the calculated
value of $\lambda$ as a function of vibrational energy $\hbar \omega$. It is found that there are only three normal modes with $\lambda > 1$,
specifically modes of $\hbar \omega = 2.0$, 2.5, 3.7 meV with $\lambda = 1.27$, 1.33, 1.46, respectively \cite{BURZ14}.
The mode ``b'' in Fig.~\ref{fig:ev}(b) is antisymmetric about the C$_2$ symmetry axis, while the mode ``c'' in Fig.~\ref{fig:ev}(c) is symmetric
about the C$_2$ symmetry axis.

\section{Model Hamiltonian and master equation}

In this section, we present the formalism to calculate transport properties from the model Hamiltonian, adapted from
Refs.~\cite{MITR04,KOCH06} to include the molecular spin Hamiltonian and the multiple vibrational modes.

\subsection{Model Hamiltonian}

We consider the following model Hamiltonian ${\cal H}={\cal H}_{\rm el} + {\cal H}_{\rm mol} + {\cal H}_{\rm t}$:
\begin{eqnarray}
{\cal H}_{\rm el} &=& \sum_{\alpha=L,R} \sum_{k,\sigma} \epsilon_{k,\sigma}^{\alpha}
a_{k,\sigma}^{\alpha \dag} a_{k,\sigma}^{\alpha}, \: \: \: \: \: \:
{\cal H}_{\rm t} = \sum_{\alpha =L,R} \sum_{k,\sigma} ( t_{\alpha}^{\star} c_{\sigma}^{\dag} a_{k,\sigma}^{\alpha}
 + t_{\alpha} a_{k,\sigma}^{\alpha \dag} c_{\sigma}), \\
 {\cal H}_{\rm mol} &=& - D_N (S_z^{(N)})^2 + (\epsilon - eV_g) \sum_{\sigma} c_{\sigma}^{\dag} c_{\sigma}
 + g \mu_B \vec{S}^{(N)} \cdot \vec{B} \nonumber \\
  & & + \sum_i \hbar \omega_i d^{\dag}_i d_i + \sum_i \lambda_i \hbar \omega_i (d^{\dag}_i + d_i)
\sum_{\sigma} c_{\sigma}^{\dag} c_{\sigma},
\label{eq:ham}
\end{eqnarray}
where $a_{k,\sigma}^{\alpha \dag}$ and $a_{k,\sigma}^{\alpha}$ are creation and annihilation operators for an electron at the electrode
$\alpha$ with energy $\epsilon_{k,\sigma}^{\alpha}$, momentum $\vec{k}$, and spin $\sigma$. Here $c_{\sigma}^{\dag}$ and $c_{\sigma}$ are
creation and annihilation operators for an electron with spin $\sigma$ at the molecular orbital $\epsilon$ or the LUMO. The parameter
$t_{\alpha}^{\star}$ in ${\cal H}_{\rm t}$ describes electron tunneling from the electrode $\alpha$ to the SMM. Symmetric tunneling is
assumed such that $t_L = t_R$. In ${\cal H}_{\rm mol}$, $D_N (> 0)$ is the uniaxial magnetic anisotropy parameter for the charge state
$N$ with the total spin $S^{(N)}$. The transverse magnetic anisotropy is neglected, since the uniaxial magnetic anisotropy 
and an applied magnetic field are much greater than the transverse anisotropy. A charging energy of the Fe$_4$ is about 2.3~eV based on
our DFT calculation, and experimental conductance maps show only two Coulomb diamonds \cite{ZYAZ10,BURZ12,BURZ14}. Therefore, we 
consider only two charge states: the neutral ($N=0$) state with $S=5$ and the singly charged ($N=1$) state with $S=9/2$. 
The second and third terms in ${\cal H}_{\rm mol}$ represent changing the orbital energy by gate voltage $V_g$ and
the Zeeman energy with $g=2$, respectively. The second line in ${\cal H}_{\rm mol}$ comprises (a) the energies of independent harmonic
oscillators with vibrational angular frequencies $\omega_i$ and (b) the coupling between electric charge and vibrational modes with
coupling strengths $\lambda_i$. Here $d_{i}^{\dag}$ and $d_{i}$ are creation and annihilation operators for the $i$-th quantized
vibrational mode or vibron. It is assumed that the vibrational frequencies are not sensitive to the charge state of the Fe$_4$.

For a weak coupling between the electrodes and the SMM, ${\cal H}_{\rm t}$ is a small perturbation to ${\cal H}_{\rm el}$ and ${\cal H}_{\rm mol}$.
Thus, a total wave function $| \Psi \rangle$ can be written as a direct product of a wave function of the electrode $\alpha$,
$|\Phi_{\alpha} \rangle$, and the molecular eigenstate $|q \rangle$. Based on the Born-Oppenheimer approximation, the latter can be given by
$| \psi^N_{m,q} \rangle \otimes | n_q \rangle$, where $| \psi^N_{m,q} \rangle$ describes an electronic charge and magnetic state and
$| n_q \rangle$ is a vibrational eigenstate of the SMM with $n_q$ vibrons. For $p$ vibrational modes, $n_q=n_1 + n_2 + ... + n_p$, where $n_i$
is a quantum number of the $i$-th vibrational mode.

When the SMM is charged, the electron-vibron coupling gives rise to off-diagonal terms in the vibrational part of the ${\cal H}_{\rm mol}$ matrix.
These terms can be eliminated by applying a canonical transformation \cite{MITR04,KOCH06} to the Hamiltonian, such as
$e^{\hat Y} \hat{O} e^{-{\hat Y}}$, where $\hat{O}$ is an observable operator and
${\hat Y} \equiv -\sum_i \lambda_i (d^{\dag}_i - d_i) \sum_{\sigma} c_{\sigma}^{\dag} c_{\sigma}$. After the transformation, the molecular
Hamiltonian becomes diagonal with respect to the new vibron creation and annihilation operators $d_i^{\prime,\dag}$ and $d_i^{\prime}$,
where $d_i^{\prime} = d_i + \lambda_i \sum_{\sigma} c^{\dag}_{\sigma} c_{\sigma}$. The canonical transformation shifts $\epsilon$ to
$\epsilon^{\prime}=\epsilon - \sum_i \lambda_i^2 \hbar \omega_i$, while $t_{\alpha}$ is modified to
$t_{\alpha} \mathrm{exp}[-\sum_i \lambda_i (d^{\dag}_i - d_i)]$. This energy shift corresponds to a shift of polaron energy caused by adjustment
of the ions following the electron tunneled to the molecule. Henceforth, we drop all primes in the operators, parameters, and Hamiltonians.

\subsection{Transition rates}

In the sequential tunneling limit [Fig.~\ref{fig:geo}(c)], we write transition rates $R_{i \rightarrow f}$ from the initial state
$|\Psi_i \rangle$ to the final state $|\Psi_f \rangle$, to the lowest order in ${\cal H}_{\rm t}$, as
\begin{eqnarray}
R_{i \rightarrow f} &=& \frac{2 \pi}{\hbar} | \langle \Psi_f |{\cal H}_{\rm t}| \Psi_i \rangle |^2
\delta (E_f - E_i),
\label{eq:Rtran} \\
{\cal H}_{\rm t} &=& \sum_{\alpha=L,R} \sum_{k,\sigma}
( t_{\alpha}^{\star} \hat{X}^{\dag} c_{\sigma}^{\dag} a_{k,\sigma}^{\alpha}
 + t_{\alpha} \hat{X} a_{k,\sigma}^{\alpha \dag} c_{\sigma} ),
 \: \: \: \: \: \hat{X} \equiv \mathrm{exp}[-\sum_i \lambda_i (d^{\dag}_i - d_i)]
\label{eq:Htran_new}
\end{eqnarray}
where $E_f$ and $E_i$ are the final and initial energies, and ${\cal H}_{\rm t}$ is the new tunneling Hamiltonian after the canonical
transformation. In these rates we integrate over degrees of freedom of the electrodes and take into account thermal distributions of the electrons
in the electrodes by the Fermi-Dirac distribution function $f(E)$. Then the transition rates can be written in terms of degrees of freedom of the
SMM only \cite{KOCH06}.

Let us first discuss transition rates $\gamma_{\alpha}^{q \rightarrow r}$ from a magnetic level in the $N=0$ state
$|q \rangle = |\psi^{N=0}_{M,q}, n_q \rangle$ to a level in the $N=1$ state $|r \rangle = |\psi^{N=1}_{m,r}, n_r \rangle$, i.e.,
electron tunneling from the electrode $\alpha$ to the SMM. The rates are given by
\begin{eqnarray}
\gamma_{\alpha}^{q \rightarrow r} &=& \sum_{\sigma} W^{\sigma, \alpha}_{q \rightarrow r} f( \bar{\epsilon} - \mu_{\alpha} )
{\cal F}_{n_q,n_r},
\label{eq:tran01}
\end{eqnarray}
where $W^{\sigma, \alpha}_{q \rightarrow r}$ and ${\cal F}_{n_q,n_r}$ represent transition rates associated with the electronic and nuclear
degrees of freedom, respectively. Here $\bar{\epsilon}$ is defined to be $\epsilon_m^{N=1} - \epsilon_M^{N=0} + (n_r - n_q) \hbar \omega$ for
a single vibrational mode, where $\epsilon_{m,M}^{N}$ contain orbital and magnetic energies of the SMM for the charge state $N$. For multiple
vibrational modes, indices for individual modes are introduced in $n_q$ and $n_r$, following the scheme in Refs.~\cite{SELD11,RUHO94,RUHO00}. The chemical
potential of the left and right electrodes are $\mu_L= -\mu_R = eV/2$, where $V$ is a bias voltage. In Eq.~(\ref{eq:tran01}),
$f(\bar{\epsilon} - \mu_{\alpha})$ is included in the transition rates since electrons tunnel from the electrode $\alpha$. We discuss the
electronic and nuclear parts of the rates separately.

The electronic part of the rates is given by
\begin{eqnarray}
W^{\sigma, \alpha}_{q \rightarrow r} &=& \frac{2 \pi}{\hbar} {\cal D}^{\alpha}_{\sigma} |t_{\alpha}|^2
| \langle \psi^{N=1}_{m,r} | c^{\dag}_{\sigma} | \psi^{N=0}_{M,q} \rangle |^2,
\label{eq:W} \\
| \psi^{N=0}_{M,q} \rangle &=& \sum_l u_l |S=5, M_l \rangle, \: \: \: \: M_l= -5, -4, ..., 4, 5,  \\
| \psi^{N=1}_{m,r} \rangle &=& \sum_j v_j |S=9/2, m_j \rangle, \: \: \: \: m_j= -9/2, -7/2, ..., 7/2, 9/2,
\end{eqnarray}
where ${\cal D}^{\alpha}_{\sigma}$ is the density of states of the electrode ${\alpha}$ near the Fermi level $E_F$, which is assumed to be
constant and is independent of $\alpha$ and $\sigma$. The initial and final electronic states of the SMM, $|\psi^{N=0}_{M,q} \rangle$
and $|\psi^{N=1}_{m,r} \rangle$, can be expressed as a linear combination of the eigenstates of $S_z$ for $S=5$ and $S=9/2$, respectively.
The matrix elements $\langle \psi^{N=1}_{m,r} | c^{\dag}_{\sigma} | \psi^{N=0}_{M,q} \rangle$ in $W^{\sigma, \alpha}_{q \rightarrow r}$
dictate selection rules such as $|M - m|=1/2$ and $\Delta N=\pm 1$, and they are evaluated by using the Clebsch-Gordon
coefficients.

The nuclear part of the rates, ${\cal F}_{n_q,n_r}$, is called the Franck-Condon factor \cite{KOCH06}, and it is symmetric with respect to
the indices. The factor is defined to be $| {\cal J}_{n_q,n_r}|^2$, where ${\cal J}_{n_q,n_r}$ is an overlap matrix between the nuclear
wave functions of the $N=0$ and $N=1$ states \cite{KOCH06,SELD08,SELD11}, i.e.,
\begin{eqnarray}
{\cal J}_{n_q,n_r} &=& \langle n_r | \hat{X} | n_q \rangle.
\label{eq:FC}
\end{eqnarray}
In the case of $p$ vibrational modes, for $n_q=n_r=0$, it is known that
\begin{eqnarray}
{\cal J}_{0,0} &=& \exp[-\sum_{k=1}^{p} \lambda_k^2/2], \: \: \: \: {\cal F}_{0,0}=\exp[-\sum_{k=1}^{p} \lambda_k^2].
\label{eq:FC00}
\end{eqnarray}
For the rest of $n_q$ and $n_r$ values, the overlap matrix elements can be found by applying the following recursion relations
\cite{RUHO94,RUHO00}:
\begin{eqnarray}
{\cal J}_{n,n^{\prime}} &=& -\frac{\lambda_i}{\sqrt{n_i}} {\cal J}_{n_i - 1} +
\frac{\sqrt{n_i^{\prime}}}{\sqrt{n_i}} {\cal J}_{n_i - 1, n_i^{\prime} - 1} \: \: \: \: \: (n_i > 0),\\
{\cal J}_{n,n^{\prime}} &=&  \frac{\lambda_i}{\sqrt{n_i^{\prime}}} {\cal J}_{n_i^{\prime} - 1} +
\frac{\sqrt{n_i}}{\sqrt{n_i^{\prime}}} {\cal J}_{n_i - 1, n_i^{\prime} - 1} \: \: \: \: \: (n_i^{\prime} > 0),
\end{eqnarray}
where $n=(n_1,...,n_p)$ and $n^{\prime}=(n_1^{\prime},...,n_p^{\prime})$. In ${\cal J}_{n_i - 1}$, the quantum number $n_i$ is lowered
by one with the rest of the quantum numbers fixed, while in ${\cal J}_{n_i - 1, n_i^{\prime} - 1}$, both quantum numbers $n_i$ and
$n_i^{\prime}$ are lowered by one with the rest fixed. For example, for a single vibrational mode, we find that
${\cal J}_{0,1}=\lambda e^{-\lambda^2/2}$ and ${\cal F}_{0,1}=\lambda^2 e^{-\lambda^2}$.

Now we discuss the transition rates $\gamma_{\alpha}^{r \rightarrow q}$ from the $N=1$ state $|r \rangle = |\psi^{N=1}_{m,r}, n_r \rangle$
to the $N=0$ state $|q \rangle = |\psi^{N=0}_{M,q}, n_q \rangle$, i.e., electron tunneling from the SMM to the electrode $\alpha$. Similarly
to Eq.~(\ref{eq:tran01}), the rates are given by
\begin{eqnarray}
\gamma_{\alpha}^{r \rightarrow q} &=& \sum_{\sigma} W^{\sigma, \alpha}_{r \rightarrow q}
[1 - f( \bar{\epsilon} - \mu_{\alpha} )] {\cal F}_{n_r,n_q},
\label{eq:W2}
\end{eqnarray}
where $1-f(\bar{\epsilon} - \mu_{\alpha})$ appears since an energy level $\bar{\epsilon} - \mu_{\alpha}$ must be unoccupied for an electron
to tunnel back to the electrode $\alpha$.

\subsection{Master equation}

A probability $P_q$ of the molecular state $|q \rangle$ being occupied, satisfies the master equation
\begin{eqnarray}
\frac{dP_q}{dt} &=& - P_q \sum_{\alpha=L,R} \sum_{r} \gamma_{\alpha}^{q \rightarrow r}
+ \sum_{\alpha=L,R} \sum_{r} \gamma_{\alpha}^{r \rightarrow q} P_r,
\label{eq:master}
\end{eqnarray}
where the summation over $r$ runs for the orbital, magnetic, and vibrational degrees of freedom. The first (second) term sums up all allowed transitions from (to) the state $| q \rangle$. We assume that the vibrons are not equilibrated, in other words, they have a long relaxation
time. For steady-state probabilities $P_q$, we solve $dP_q/dt = 0$ by applying the bi-conjugate gradient stabilized method \cite{SELD11,VORS92}. Starting with the Boltzmann distribution at $V=0$ as initial probabilities, we achieve a fast convergence to the steady-state solution for
non-zero bias voltages. Finally, we compute the current $I_{\alpha}$ from the electrode $\alpha$ to the SMM using the steady-state
probabilities and transition rates,
\begin{eqnarray}
I_{\alpha=L,R} &=& e \sum_{q,r} \gamma_{\alpha}^{|N=0,q \rangle \rightarrow |N=1,r \rangle} P_q
- e \sum_{q,r} \gamma_{\alpha}^{|N=1,r \rangle \rightarrow |N=0,q \rangle} P_r,
\end{eqnarray}
where the sums over $q$ and $r$ run for all the orbital, magnetic, and vibrational indices. In our set-up, the current is positive
when an electron tunnels from the left electrode to the SMM (or from the SMM to the right electrode), while it is negative when an
electron tunnels from the SMM to the left electrode (or from the right electrode to the SMM). The total current $I=(I_L-I_R)/2$. For
symmetric coupling to the electrodes, we have that $I_L=-I_R$. A differential conductance $dI/dV$ is computed numerically from
current-voltage ($I-V$) characteristics by using a small bias interval of $\Delta V=0.01$ or 0.05~mV.

\section{Results and Discussion: Transport Properties}

We present the $I-V$ characteristics and $dI/dV$ vs $V$ as a function of $V_g$, temperature $T$, and applied $B$ field, obtained by solving the
master equation Eq.~(\ref{eq:master}) with the DFT-calculated parameter values. We use $D_{N=0}=0.056$~meV and $D_{N=1}=0.062$~meV. The
value of $D_{N=1}$ is chosen to be 10\% greater than the value of $D_{N=0}$, which is consistent with the experimental data \cite{BURZ12}.
We consider up to 9 vibrons ($n=9$), which is large enough that the transport
properties do not change with a further increase of $n$ in the ranges of $V$ and $V_g$ of interest. The level broadening
$\Gamma=2 \pi {\cal D}|t|^2$ is taken as 0.01~meV, which satisfies that $\Gamma \ll k_B T$, $\hbar \omega$. In the sequential tunneling
limit, the $\Gamma$ value plays a role of units in the current and conductance.

Regarding the electron-vibron coupling, we consider two cases: (i) a single vibrational mode with $\lambda > 1$, such as
$\hbar \omega=2.0$~meV with $\lambda=1.27$ [Fig.~\ref{fig:ev}(c)], and (ii) three vibrational modes with $\lambda > 1$, such as
$\hbar \omega_{1,2,3} = 2.0$, 2.5, 3.7 meV with $\lambda_{1,2,3} = 1.27$, 1.33, 1.46 [inset of Fig.~\ref{fig:ev}(a)], which are only
modes with $\lambda > 1$ from the DFT calculation (Sec.III.B). The case (i) is an instructive example of the electron-vibron
coupling. The case (ii) approximates to the case that all of the vibrational modes are included in
${\cal H}_{\rm mol}$, Eq.~(\ref{eq:ham}), since the modes with $\lambda < 1$ would not significantly contribute to the sequential
tunneling at low bias. This is justified because of their exponential contributions to the Franck-Condon factor, Eq.~(\ref{eq:FC00}). We
also confirm that this is the case from actual calculations of the $I-V$ and $dI/dV$ with an additional low-$\lambda$ normal mode to the
case (ii). We first present the basic features and magnetic-field dependencies of the conductance peaks for the case (i) and then those
for the case (ii).

\subsection{Case (i): Basic features}


\begin{figure}[h]
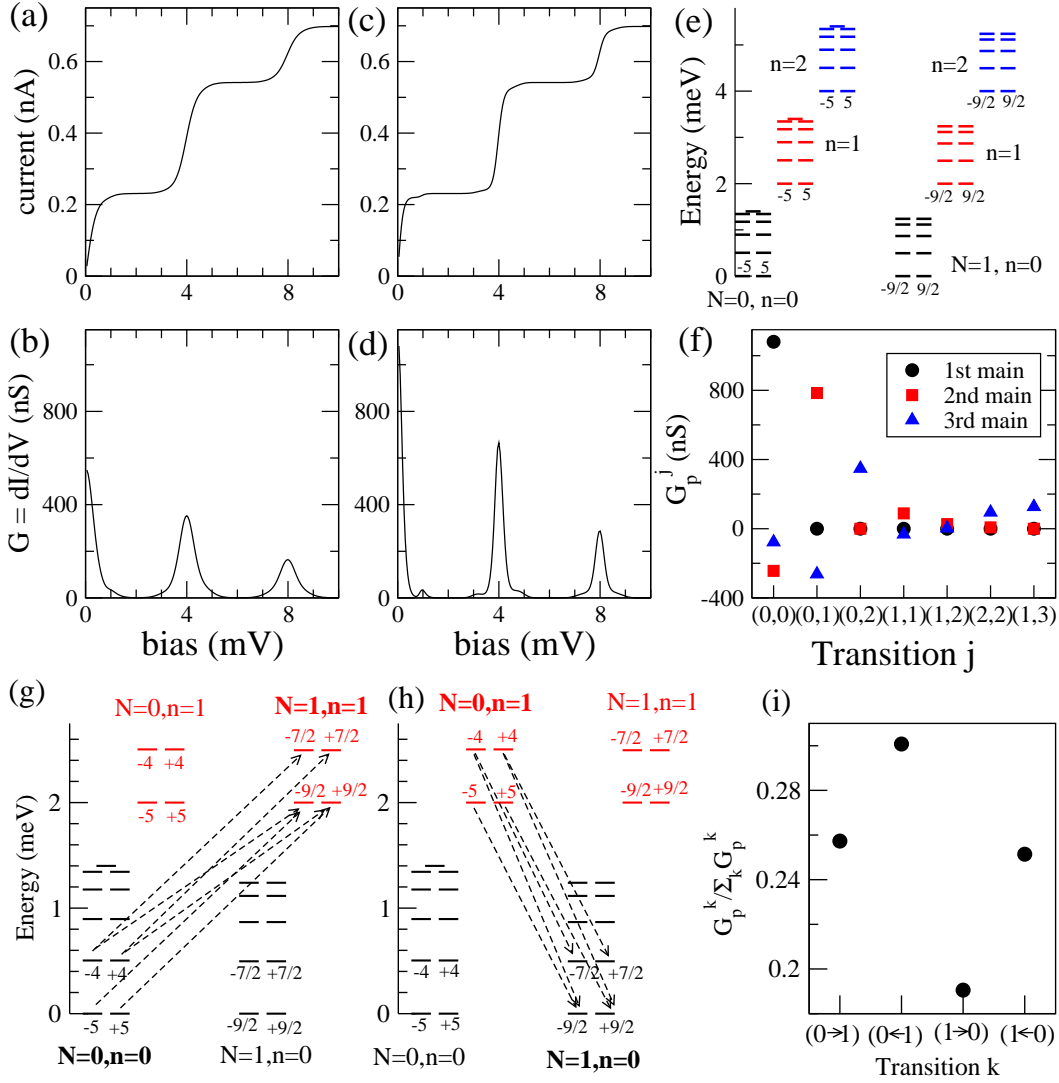

\includegraphics[width=0.85\textwidth]{nmodes_1.eps}
\includegraphics[width=0.85\textwidth]{B_zero_supp_v02.eps}
\caption{(Color Online) Calculated $I-V$ and $dI/dV$ vs $V$ at the charge degeneracy point for the case (i) at
$T=1.16$~K [(a),(b)] and $T=0.58$~K [(c),(d)]. (e) Magnetic energy levels in the vibrational $n=0$, $n=1$, and $n=2$ states
for the two charge states $N=0$ and $N=1$. For each set of the magnetic levels, the left column, the center, and the right column 
correspond to the levels $M < 0$, $M=0$, and $M > 0$, respectively. (f) Contributions of different transitions $j$ to the first 
(leftmost), second, and third main $dI/dV$ peak heights in (d). See the main text for the definitions of the transitions 
($n$,$n^{\prime}$). (g) and (h) Dominant transition pathways for the transitions (0$\rightarrow$1) and (1$\rightarrow$0). 
(i) Contributions of transitions $k$ within the transitions ($n=0$,$n^{\prime}=1$) to the height of the second main peak 
in (d).}
\label{fig:Tdep_1}
\end{figure}

\begin{figure}
\includegraphics[width=0.85\textwidth]{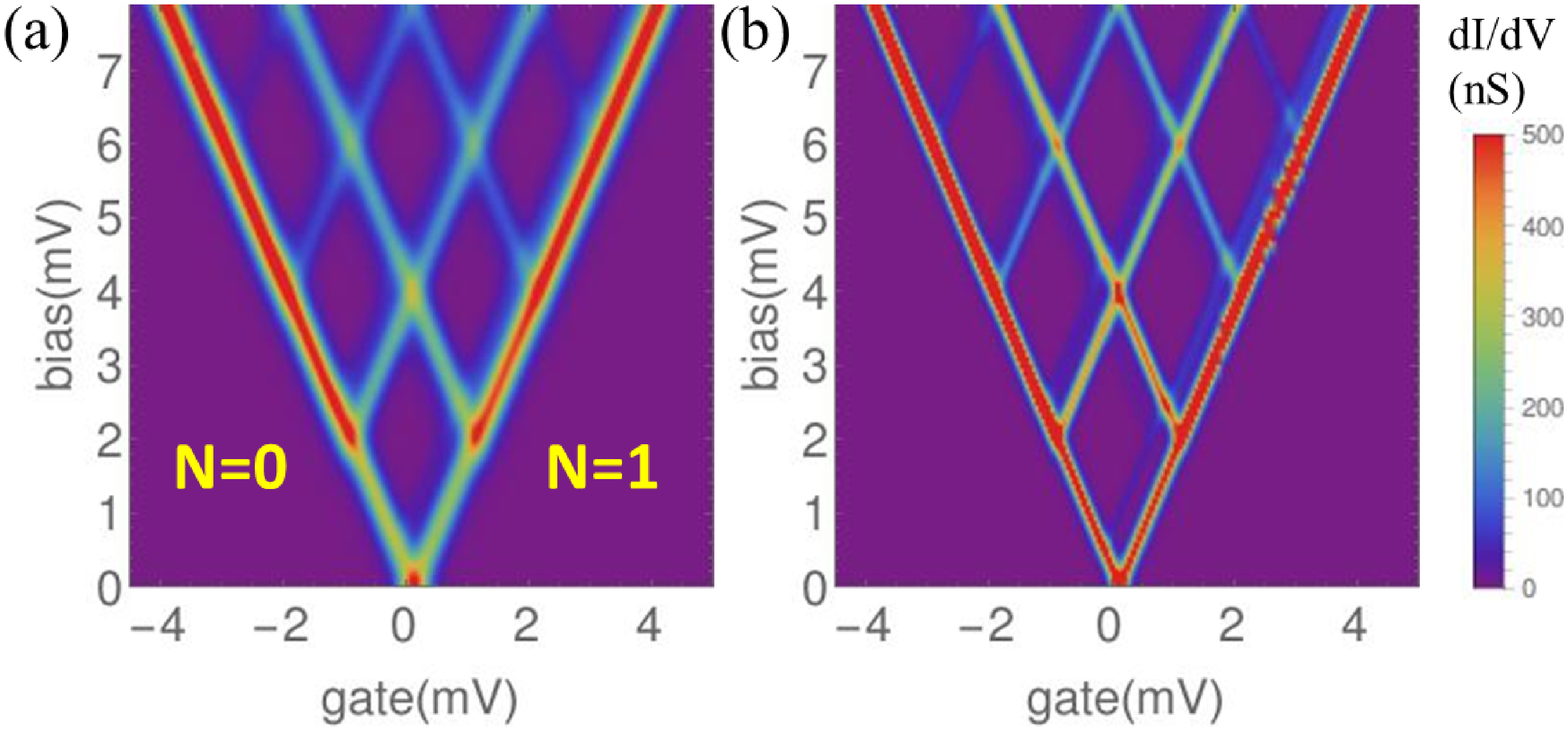}
\caption{(Color Online) Calculated $dI/dV$ maps as a function of $V$ and $V_g$ for the case (i) at $T=1.16$~K (a) and $T=0.58$~K (b).}
\label{fig:CB_1}
\end{figure}

Figures~\ref{fig:Tdep_1}(a)-(d) show the $I-V$ curve and $dI/dV$ vs $V$ for the case (i) at $T=1.16$~K and 0.58~K ($\sim 0.05$~meV),
for the gate voltage where the lowest magnetic levels of the $N=0$ and $N=1$ charge states are degenerate at zero bias, i.e.,
charge degeneracy point. This gate voltage is set to zero. The steps in the current and the $dI/dV$ peaks appear at $V = 2 n \hbar \omega$ ($n=0$,1,2,...), where the factor of 2 is due to the symmetric bias application [Fig.~\ref{fig:geo}(c)]. The first
peak at $V=0$ arises from the vibrational ground state ($n=0$), while the second and third peaks at $V=4.0$ and 8.0~mV come from vibrational
excitations ($n=1$ and $n=2$), respectively. Figures~\ref{fig:CB_1}(a) and (b) reveal $dI/dV$ ($=G$) maps as a function of $V$ and $V_g$, i.e.,
stability diagrams, at 1.16~K and 0.58~K, respectively. Here the Coulomb diamond edges arise from the sequential tunneling via the lowest
doublets in the $n=0$ state, while the evenly spaced peaks parallel to the Coulomb diamond edges originate from the vibrational excitations.
As $T$ is lowered, overall features of the peaks do not change, while the peaks become sharper with more apparent fine structures.

We now analyze the heights of the $dI/dV$ peaks at 0.58~K in detail [Fig.~\ref{fig:Tdep_1}(d)]. The $dI/dV$ peak height decreases as $n$
increases. This implies that the sequential tunneling via the vibrational ground states is dominant over the tunneling via the vibrational
excitations for $\lambda=1.27$. This feature qualitatively differs from the case (ii) (Sec.V.C). A peak height at a fixed temperature is
determined by the Franck-Condon factor, the electronic part of the transition rates, and the occupation probabilities. We introduce
simplified notations for transitions between the $N=0$ and $N=1$ states: ($n$,$n^{\prime}$)$\equiv$
$\{ |\psi^{N=0}_{M} \rangle \otimes |n \rangle \rightarrow |\psi^{N=1}_{m} \rangle \otimes |n^{\prime} \rangle$,
$|\psi^{N=0}_{M} \rangle \otimes |n \rangle \leftarrow |\psi^{N=1}_{m} \rangle \otimes |n^{\prime} \rangle $,
$|\psi^{N=0}_{M} \rangle \otimes |n^{\prime} \rangle \rightarrow |\psi^{N=1}_{m} \rangle \otimes |n \rangle$,
$|\psi^{N=0}_{M} \rangle \otimes |n^{\prime} \rangle \leftarrow |\psi^{N=1}_{m} \rangle \otimes |n \rangle \}$
$\equiv$ $\{ (n \rightarrow n^{\prime}),(n \leftarrow n^{\prime}),(n^{\prime} \rightarrow n),(n^{\prime} \leftarrow n) \}$.
Here ($n$,$n^{\prime}$) contain all possible tunneling paths including all magnetic levels allowed by the selection rules.
Several values of the Franck-Condon factor for ($n$,$n^{\prime}$), are listed in Table~\ref{table3} in the Appendix.

Figure~\ref{fig:Tdep_1}(f) shows contributions of different transitions ($n$,$n^{\prime}$) to the first, second, and third peak heights, $G_p^{(n,n^{\prime})}$. For the first peak height, only transitions ($n=0,n^{\prime}=0$) contribute. Resonant tunneling occurs
via the lowest doublets ($M=\pm 5$ and $m=\pm 9/2$) in the $n=0$ state because they are only occupied levels at 0.58~K. The zero-field
splitting is one order of magnitude larger than the thermal energy, and so levels other than the doublets are not occupied [Figs.~\ref{fig:geo}(b),\ref{fig:Tdep_1}(e)].

Regarding the second peak height, transitions ($n=0$,$n^{\prime}=1$) dominantly contribute, while transitions ($n=1$,$n^{\prime}=1$) slightly
involve in the tunneling [Fig.~\ref{fig:Tdep_1}(f)]. In this case, all the levels in the $n=0$ state and some low-lying levels in the $n=1$
are occupied. At $V=4.0$~mV, the transitions ($n=0$,$n^{\prime}=0$) lower the second peak height because the occupation probabilities
of the levels in the $n=0$ state differ from those in the case of zero bias. When all the contributions are summed, the second peak
is found to have a smaller height than the first peak. Let us discuss in detail the tunneling via ($n=0$,$n^{\prime}=1$) at $V=4.0$~mV. The contributions of ($n=0$,$n^{\prime}=1$) can be decomposed into those of ($0 \rightarrow 1$), ($0 \leftarrow 1$), ($1 \rightarrow 0$), and
($1 \leftarrow 0$), as shown in Fig.~\ref{fig:Tdep_1}(i). The transition ($0 \leftarrow 1$) gives the highest peak value $G_p$ among the four
transitions. In the case of ($0 \rightarrow 1$), as shown in Fig.~\ref{fig:Tdep_1}(g), each of the levels $M=\pm 4, \pm 3, \pm 2, \pm 1, 0$
in the $n=0$ state can tunnel to two $m$ levels in the $n=1$ state, such as $M=-4$ in the $n=0$ state to $m=-7/2, -9/2$ in the
$n=1$ state, but the lowest level $M=5$ ($M=-5$) in the $n=0$ state can transit only to one $m$ level in the $n=1$ state such as $m=9/2$
($m=-9/2$). However, for the reverse transition, ($0 \leftarrow 1$), each of all levels in the $n=1$ state can tunnel to two $M$ levels in
the $n=0$ state. In addition, the separation between the level $m=-9/2$ in the $n=1$ state and the level $M=-4$ in the $n=0$ state is
$\hbar \omega - 9 D_0$ which is less than $\hbar \omega$. These two factors are the reasons that the contribution of ($0 \leftarrow 1$) to the
$G_p$ is higher than that of ($0 \rightarrow 1$). An interesting case is the transition ($1 \rightarrow 0$) shown
in Fig.~\ref{fig:Tdep_1}(h). In this case, two of the allowed transitions require a higher bias voltage than 4.0~mV. The energy difference
between the level $M=-4$ ($M=4$) in the $n=1$ state and the level $m=-9/2$ ($m=9/2$) in the $n=0$ state is $\hbar \omega + 9 D_0$. This
energy difference prevents the levels $M=\pm 4$ in the $n=1$ state from being significantly occupied at $eV/2 = \hbar \omega$. As a consequence,
the transition ($1 \rightarrow 0$) participates in the tunneling much less than the other three transitions, as confirmed in
Fig.~\ref{fig:Tdep_1}(i).

For the third peak height, transitions ($n=0$,$n^{\prime}=2$) play a leading role, with considerable contributions of transitions ($n=1$,$n^{\prime}=3$) and ($n=2$,$n^{\prime}=2$) [Fig.~\ref{fig:Tdep_1}(f)]. At $V=8.0$~mV, all the levels in the $n=0$ and $n=1$ states as well
as some low-lying levels in the $n=2$ and $n=3$ states are involved in the tunneling. The occupation of the levels in the $n=2$ and $n=3$ 
states significantly modifies the occupation of the levels in $n=0$ and $n=1$ states compared to the case of $V=4.0$~mV. Accordingly,
this modification causes the transitions ($n=0$,$n^{\prime}=0$) and ($n=0$,$n^{\prime}=1$) to contribute to the third peak height less 
than in the case of $V=4.0$~mV. Overall, when all the contributions are added, the third peak has a smaller height than the second peak.

We now examine the magnetic anisotropy effect on the $dI/dV$ map at 0.58~K, as shown in Figs.~\ref{fig:Tdep_1}(d) and \ref{fig:CB_1}(b).
The small (or satellite) peak at 1.0~mV and the flat shoulders around the second and third main peaks in Fig.~\ref{fig:Tdep_1}(d), are
signatures of the magnetic anisotropy. Since the zero-field splitting (0.5~meV) is a maximum energy difference between adjacent levels
for a given $N$ and $n$ state, at a bias voltage of 1.0~mV, all $M$ and $m$ levels in the $n=0$ state are accessible. Thus, all the
levels in the $n=0$ state are equally occupied and they contribute to the satellite peak at 1.0~mV. Additional satellite peaks
are not found despite increasing a bias voltage, until some low-lying levels in the $n=1$ state become occupied. The left-hand (right-hand)
shoulder of the second main peak in Fig.~\ref{fig:Tdep_1}(d) is attributed to tunneling to the lowest doublet in the $n=1$ ($n=2$) state
barely occupied.

\subsection{Case (i): Magnetic field dependence}

\begin{figure}[h]
\includegraphics[width=0.85\textwidth]{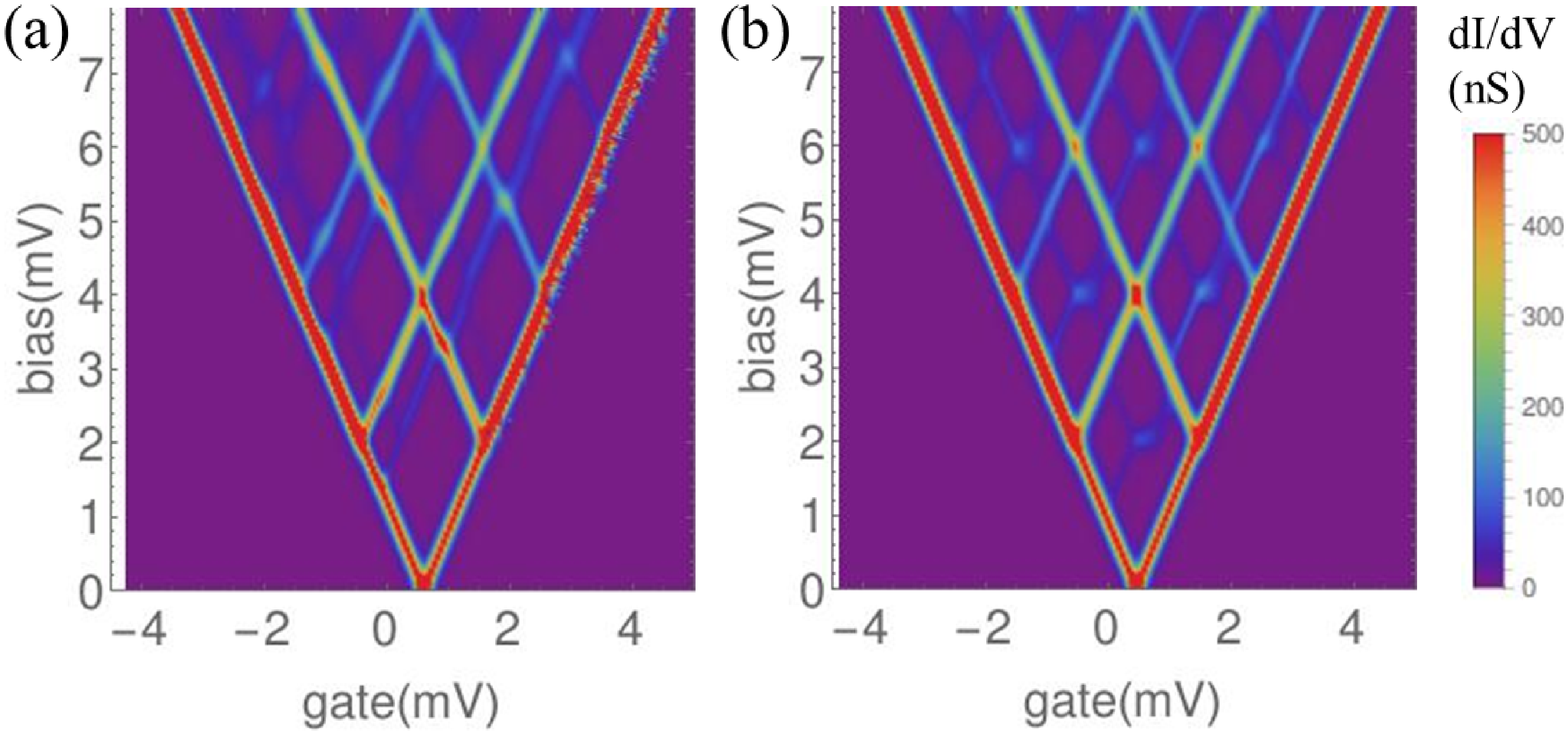}
\includegraphics[width=0.85\textwidth]{nmodes_1_B8T.eps}
\caption{(Color online) Calculated $dI/dV$ values as a function of $V$ and $V_g$ for the case (i) at $T=0.58$~K for $B_z=8$~T (a) and
$B_x=8$~T (b). (c) and (d) Computed $dI/dV$ vs $V$ at the charge degeneracy point in (a) and (b), respectively.}
\label{fig:CB_B}
\end{figure}

Figures~\ref{fig:CB_B}(a)-(b) are stability diagrams for the case (i) at 0.58~K for $B_z=8.0$~T and $B_x=8.0$~T, respectively. The zero-bias
charge degeneracy for $B_z=8.0$~T and $B_x=8.0$~T occurs at the gate voltage of 0.61 and 0.46 mV, respectively, due to the Zeeman energy.
With an external $B$ field, it is found that the Coulomb diamonds are simply horizontally shifted from the zero $B$-field case,
in other words, that the positions of the {\it main} $dI/dV$ peaks remain the same relative to the charge degeneracy point. Compare
Figs.~\ref{fig:CB_B}(a)-(b) with Fig.~\ref{fig:CB_1}(b). Figures~\ref{fig:CB_B}(c)-(d) exhibit the $dI/dV$ vs $V$ at the charge degeneracy
point for $B_z=8.0$~T and $B_x=8.0$~T, respectively. Compare Figs.~\ref{fig:CB_B}(c)-(d) with Fig.~\ref{fig:Tdep_1}(d). The shift of the
main peaks was observed in experiment \cite{BURZ14}, and it is consistent with the vibrational origin of the main peaks.

A further comparison between Figs.~\ref{fig:CB_B}(c)-(d) with Fig.~\ref{fig:Tdep_1}(d) reveals two interesting aspects of the $B$-field
dependence of the peaks: (1) The heights of the main peaks are greatly modified with the direction as well as magnitude
of applied $B$ field, which is a signature of the magnetic anisotropy; (2) Both the positions and the heights of the satellite peaks strongly
depend on the direction and magnitude of applied $B$ field. Note that the two effects are found because the magnetic anisotropy
barrier or the zero-field splitting is on the same order of magnitude as the vibrational energy. In this section, we present features of
the main peak height for the case (i) as a function of $B$ field followed by those of the positions and the heights of the satellite peaks,
by considering two $B$-field orientations ($z$ and $x$ axes) for $0 \le B \le 24.0$~T. Our calculations are
carried out at 0.58~K and at a gate voltage corresponding to the charge degeneracy point for each $B$-field value.

\subsubsection{$B_z$-field dependence of main peaks}


\begin{figure}[h]
\includegraphics[width=0.85\textwidth]{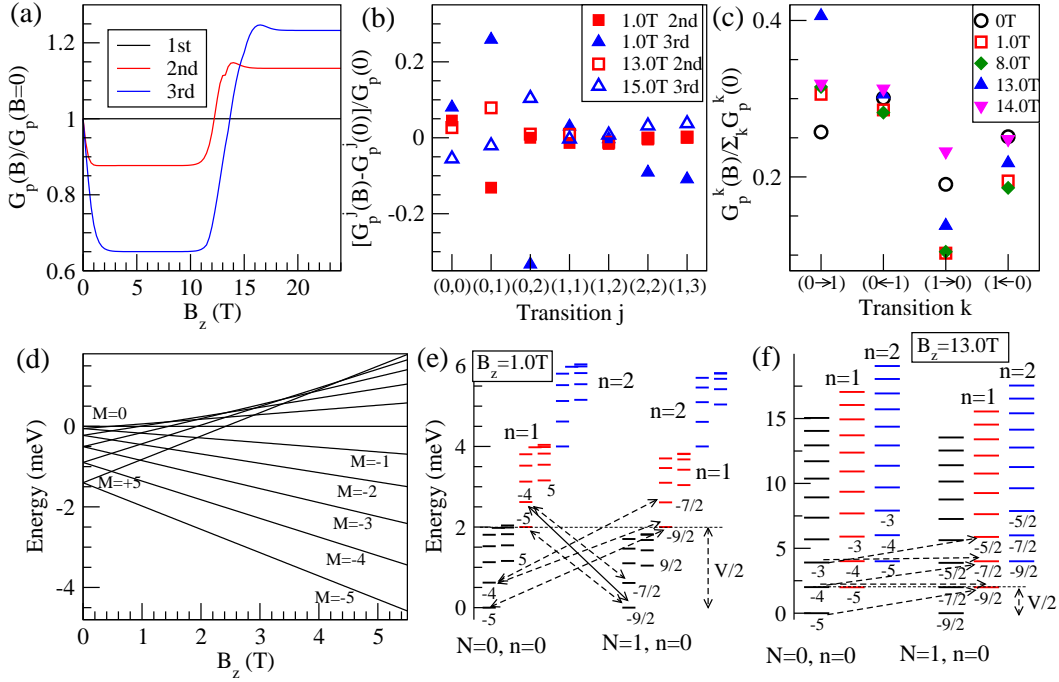}
\caption{(Color online) (a) Ratio of the heights of the first (black), second (red), and third (blue) main $G=dI/dV$ peaks at $B_z$ to
the heights of the corresponding $B=0$ peaks, $G_p(B)/G_p(0)$, vs $B_z$ for the case (i) at 0.58~K.
(b) Changes of contributions of different transitions $j$
to the second and third main peaks $[G_p^{j}(B) - G_p^{j}(B=0)]$, relative to the zero $B$ case, normalized by the height of the second and
third main peaks at zero $B$, $G_p(0)$, computed at $B_z=1.0$, 13.0, and 15.0~T. (c) Contributions of transitions $k$ within the transitions
($n=0$,$n^{\prime}=1$) to the second peak at $B_z=0, 1.0, 8.0, 13.0$, and 14.0~T. (d) Evolution of the magnetic levels of $S=5$ with $B_z$.
(e) Dominant transition pathways for the second peak ($V/2=2.0$~mV) at $B_z=1.0$~T, where the solid arrow indicates the pathway
critical to the abrupt reduction of the second peak height in (a)-(c). (f) Dominant transition pathways for the second peak at
$B_z=13.0$~T dictating the sudden jump in the second peak height. In (e) and (f), the vertical arrows represent half of the peak bias
voltage.}
\label{fig:Bfield1}
\end{figure}

Figure~\ref{fig:Bfield1}(a) shows a ratio of the peak height $G_p$ at $B_z \neq 0$ to that at zero $B$, $G_p(B)/G_p(0)$, as a function of $B_z$,
for the first, second, and third main peaks. The heights of the second and third main peaks decrease abruptly at low $B$ and they remain
unchanged until about 12.0~T, above which there appear large steep rises in the heights. The effect of $B_z$ is greater on the ratio of the
third peak height than on the ratio of the second peak height. However, the first peak height does not change with $B_z$ field because
only the lowest levels ($M=-5$, $m=-9/2$) in the $n=0$ state participate in the tunneling even for $B_z \neq 0$.

Firstly, we study the $B_z$-field dependence of the second peak height by understanding how the contributions of transitions $j$ to the 
height are modified with $B_z$ relative to the $B=0$ case, i.e., by computing $[G_p^{j}(B) - G_p^{j}(B=0)]/G_p(0)$, where $j=(n,n^{\prime})$, as
shown in Fig.~\ref{fig:Bfield1}(b). It is found that the sharp decrease of the height at low $B$ ($\sim$1.0~T) is mainly caused by a
large decrease of the transitions ($n=0,n^{\prime}=1$) at $V=4.0$~mV. For further analysis, we compute $G_p^{k}(B)/\sum_k G_p^k(0)$ at
several $B_z$ values, where $k=$($0 \rightarrow 1$), ($0 \leftarrow 1$), ($1 \rightarrow 0$), ($1 \leftarrow 0$). As shown in Fig.~\ref{fig:Bfield1}(c), the large decrease of the transitions ($n=0,n^{\prime}=1$) at low $B_z$ is attributed to a large decrease
of the transition ($1 \rightarrow 0$) compared to the $B=0$ case. This decrease can be understood by examining the evolution of the
magnetic levels with $B_z$. At zero $B$, within $V=4.0$~mV, several low-lying levels, such as $M=\pm 5$, $\pm 4$ in the $n=0$ ($n=1$)
state and $m=\pm 9/2$, $\pm 7/2$ in the $n=1$ ($n=0$) state, dominantly participate in the transitions ($n=0,n^{\prime}=1$). As $B_z > 0$
increases, the $M,m < 0$ levels are shifted down, while the $M,m > 0$ levels are lifted up in energy [Fig.~\ref{fig:Bfield1}(d)].
At $B$ field somewhat above $B_z=D_1/g\mu_B=0.54$~T, the $M=5$, 4 and $m=9/2$, 7/2 levels are located quite above the $M=-5$, $-4$ and
$m=-9/2$, $-7/2$ levels. Hence, within the bias window, the $M=-5$, $-4$ levels in the $n=0$ ($n=1$) state and the $m=-9/2$, $-7/2$
levels in the $n=1$ ($n=0$) state dominantly contribute to the transitions ($n=0,n^{\prime}=1$), as shown in Fig.~\ref{fig:Bfield1}(e).
The separation between the $M=-4$ level in the $n=1$ state and the $m=-9/2$ level in the $n=0$ state, equals
$\hbar \omega + (g\mu_B B_z + 9 D_0)$. Since this separation is greater than $V/2$, the occupation of the $M=-4$ level in the $n=1$ state
decreases, and the transition ($1 \rightarrow 0$) also decreases. As a result, the transition ($n=0,n^{\prime}=1$)
considerably decreases, which leads to the drop of the second peak height at low $B$ ($\sim$1.0~T).
As $B_z$ field increases beyond 1.0~T, the separation between the two lowest levels in a given $N$ and $n$ state grows. Considering the
occupation probabilities and the transition rates, within $V=4.0$~mV, the contributions of transitions
($0 \rightarrow 1$), ($0 \leftarrow 1$), ($1 \rightarrow 0$), and ($1 \leftarrow 0$) remain almost the same as the case of $B_z=1.0$~T
[Fig.~\ref{fig:Bfield1}(c)]. Thus, the second peak height does not decrease beyond $B_z=1.0$~T. However, the situation dramatically
changes when the $B_z$ field is high enough that the spacing between the two lowest levels for a given $N$ and $n$ state equals $\hbar \omega$
[Fig.~\ref{fig:Bfield1}(f)]. This occurs at $B=(\hbar \omega - 9 D_0)/(g\mu_B)$ which is 12.9~T. In this case, the $M=-4$ ($m=-7/2$) level
in the $n=0$ state is degenerate with the $M=-5$ ($m=-9/2$) level in the $n=1$ state. Thus, at $V=2 \hbar \omega$, the occupation of the
six levels within the bias window increases compared to the case of lower $B_z$ fields, which results in an increase of the transition 
($0 \rightarrow 1$). Dominant tunneling pathways are indicated in
Fig.~\ref{fig:Bfield1}(f). More contributions from the transition ($0 \rightarrow 1$) lead to a large increase of the transitions
($n=0,n^{\prime}=1$) at high $B$ fields ($> 12.0$~T). Therefore, the peak height sharply rises above 12.0~T.

Secondly, we examine the height of the third peak. Figure~\ref{fig:Bfield1}(b) reveals that within $V=8.0$~mV, at low $B_z$, there appear
a large decrease of transitions ($n=0$,$n^{\prime}=2$) and a small decrease of transitions ($n=1$,$n^{\prime}=3$) and ($n=2$,$n^{\prime}=2$),
despite an increase of ($n=0,n^{\prime}=1$). The overall height is governed by the transitions ($n=0$,$n^{\prime}=2$). Similarly to the
second peak height, ($n=0$,$n^{\prime}=2$) can be decomposed into four sets such as ($0 \rightarrow 2$), ($0 \leftarrow 2$), ($2 \rightarrow 0$),
and ($2 \leftarrow 0$). The trend of the contribution of each of the four sets is similar to the case of the second peak if the $n=1$ state is
replaced by the $n=2$ state in the explanation. At low $B_z$, the lift of the degeneracy in the low-lying levels above $B_z=$0.54~T
drives a large reduction of the transition ($2 \rightarrow 0$), which results in the rapid drop in the peak height. The peak height
does not change beyond $B_z=2.0$~T, until the $B_z$ field is increased to the field where the spacing between the two lowest levels for a given
$N$ and $n$ state is comparable to $\hbar \omega$, similarly to the second peak. At this $B$ field (12.9~T), the second excited level in
the $n=0$ state ($M=-3$ or $m=-5/2$) and the first excited level in the $n=1$ state ($M=-4$ or $m=-7/2$) are almost degenerate with the lowest
level in the $n=2$ state ($M=-5$ or $m=-9/2$) [Fig.~\ref{fig:Bfield1}(f)]. Hence, at $V=8.0$~mV, the occupation of the levels within the bias
window substantially increases, which gives rise to a significant increase of the transition ($0 \rightarrow 2$) compared to the zero $B$-field
and low $B$ cases, in other words, an increase of the transitions ($n=0$,$n^{\prime}=2$) [Fig.~\ref{fig:Bfield1}(b)]. Consequently, the height
of the third peak sharply rises with $B_z$ field before its saturation.

\subsubsection{$B_x$-field dependence of main peaks}

\begin{figure}[h]
\includegraphics[width=0.85\textwidth]{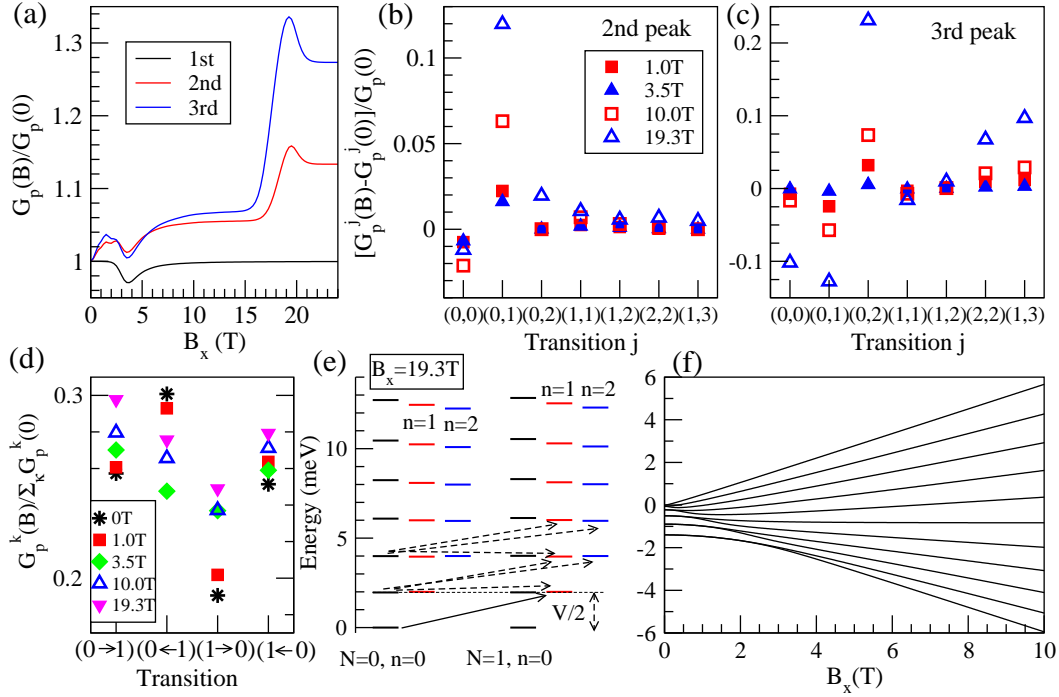}
\caption{(Color online) (a) Ratio $G_p(B)/G_p(B=0)$ for the first, second, and third main peaks as a function of $B_x$ for the case (i)
at 0.58~K. (b) and (c) $[G_p^{j}(B) - G_p^{j}(B=0)]/G_p(0)$ vs transitions $j$ for the second and third peaks at 1.0, 3.5, 10.0, and 19.3~T.
(d) Contributions of transitions $k$ within the transitions ($n=0$,$n^{\prime}=1$) to the second peak at $B_x=0, 1.0, 3.5, 10.0$, and 19.3~T.
(e) Dominant transition pathways for the second peak ($V/2=2.0$~mV) at $B_z=19.3$~T, where the solid arrow indicates the pathway
contributing most to the abrupt increase of the second peak height in (a) and (b). (f) Evolution of the magnetic levels of $S=5$ with
$B_x$.}
\label{fig:Bfield1x}
\end{figure}

Figure~\ref{fig:Bfield1x}(a) shows the ratio $G_p(B)/G_p(B=0)$ as a function of $B_x$, for the first, second, and third main peaks. As $B_x$ field
increases, the first peak height slightly decreases at low $B$ (2.0-3.5~T) and it returns to the value $G_p(0)$. The heights of the
second and third main peaks have a complex $B_x$ dependence. The heights initially increase somewhat and they slightly decrease at low $B_x$
(2.0-3.5~T). Then they gradually increase and jump up from $\sim$17.0~T. After reaching maxima near 19.3~T, the heights slightly go down before saturation. The $B_x$-field dependence qualitatively differs from the $B_z$-field dependence, which is due to the magnetic anisotropy.
Compare Fig.~\ref{fig:Bfield1x}(a) with Fig.~\ref{fig:Bfield1}(a).

Firstly, we discuss the first peak height. With a $B_x$ field, the magnetic eigenstates are admixtures of different $M_l$ levels ($m_l$ levels)
for $N=0$ state ($N=1$ state), where $M_l$ and $m_l$ are the eigenstates of $S_z$. In contrast to the case of $B_z$ field, for small $B_x$ 
fields, several
low-lying levels for a given $N$ and $n$ state remain degenerate within the thermal energy, $k_B T=$0.05~meV ($\sim$0.58~K) [Fig.~\ref{fig:Bfield1x}(f)]. For example, around $B_x=1.0$~T (2.0~T), there are three (two) low-lying doublets for a given $N$ and $n$ state [Fig.~\ref{fig:Bfield1x}(f)]. However, when the $B_x$ field increases above 3.0~T, the degeneracy of all the levels is lifted, and the separation between the adjacent levels grows with $B_x$. At $V=0$, for zero $B$, only the lowest doublet in the $N$ and $n=0$ state participate in the
tunneling, while for $B_x=$2.0-3.0~T, the first-excited level in the $N$ and $n=0$ state slightly contributes to the peak, which causes the small
decrease of the peak height. When the first-excited level is well separated from the lowest level in the $n=0$ state at higher $B_x$ fields,
the peak height resumes to the $G_p(0)$ value.

Secondly, let us examine the second peak height by computing $[G_p^j(B)-G_p^j(0)]/G_p(0)$, where $j=$($n$,$n^{\prime}$), as shown in Fig.~\ref{fig:Bfield1x}(b). It is found that the $B_x$-field dependence of the peak height is mainly determined by a $B_x$-field dependence of the transitions ($n=0$,$n^{\prime}=1$). At low $B_x$ ($\sim$1.0~T), for $V=4.0$~mV, the transition (0$\leftarrow$1) is dominant among the four possible transitions within ($n=0$,$n^{\prime}=1$) [Fig.~\ref{fig:Bfield1x}(d)]. This feature is similar to the zero $B$-field case since there are still
a few degenerate pairs for a given $N$ and $n$ state within $V=4.0$~mV. As $B_x$ increases above 3.5~T, the degeneracy of the levels is completely lifted, and so the transition (0$\leftarrow$1) used to dominantly contributing to the peak in zero $B$ field now decreases. However, the strong
mixing between different $M_l$ or $m_l$ levels in the eigenstates open up more tunneling pathways within the bias window than the $B_z$ case.
As a result, the transition (1$\rightarrow$0) used to be suppressed at zero $B$ field and $B_z$ fields now increases [Fig.~\ref{fig:Bfield1x}(d)].
With a further increase of $B_x$, the three transitions other than (1$\rightarrow$0) increase, and so the peak height goes
up. As $B_x$ field increases above 17.0~T, the spacing between the first-excited and the lowest levels for a given $N$ and $n$ state becomes
close to $\hbar \omega$, which creates more tunneling paths. For example, at 19.3~T, the first-excited level in the $n=0$ state has the same energy
as the lowest level in the $n=1$ state, and so the occupation of the levels within the bias window [Fig.~\ref{fig:Bfield1x}(e)] is higher than
the lower $B_x$-field case. At $V=4.0$~mV, the increase of the occupation makes the transitions ($n=0$,$n^{\prime}=1$) contribute more to the
peak height and it also allows the transitions ($n=0$,$n^{\prime}=2$) to participate in the tunneling. Thus, the second peak height reaches
the maximum. Some dominant tunneling pathways are indicated in Fig.~\ref{fig:Bfield1x}(e).

Thirdly, we examine the height of the third peak. The $B_x$-field dependence of the peak height dominantly arises from a $B_x$-field dependence of
the transitions ($n=0$,$n^{\prime}=2$), as shown in Fig.~\ref{fig:Bfield1x}(c). Within $V=8.0$~mV, at $B_x=$~3.5~T, the transitions (0$\leftarrow$2)
and (2$\leftarrow$0) decrease as much as transitions (0$\rightarrow$2) and (2$\rightarrow$0) increase, among the transitions
($n=0$,$n^{\prime}=2$), so that the peak height is close to the $G_p(0)$ value. As $B_x$ field increases further, the low-lying levels in the $n=0$ 
state become close to the low-lying levels in the $n=2$ state. At 19.3~T, the second-excited level in the $n=0$ state and the first-excited level 
in the $n=1$ state are almost degenerate with the lowest level in the $n=2$ state [Fig.~\ref{fig:Bfield1x}(e)]. Thus, for $V=8.0$~mV, the 
increase in the occupation of the levels within the bias window greatly enhances the tunneling via the transitions
($n=0$,$n^{\prime}=2$) and somewhat increases the transitions ($n=1$,$n^{\prime}=3$) and ($n=2$,$n^{\prime}=2$). The overall peak height becomes
the maximum despite a decrease of the transitions ($n=0$,$n^{\prime}=0$) and ($n=0$,$n^{\prime}=1$). Small contributions of new transitions
($n=0$,$n^{\prime}=3$), ($n=1$,$n^{\prime}=4$), ($n=3$,$n^{\prime}=4$) to the third peak at 19.3~T are reduced when $B_x$ field further increases.

\subsubsection{$B_z$-field dependence of satellite peaks}

\begin{figure}[h]
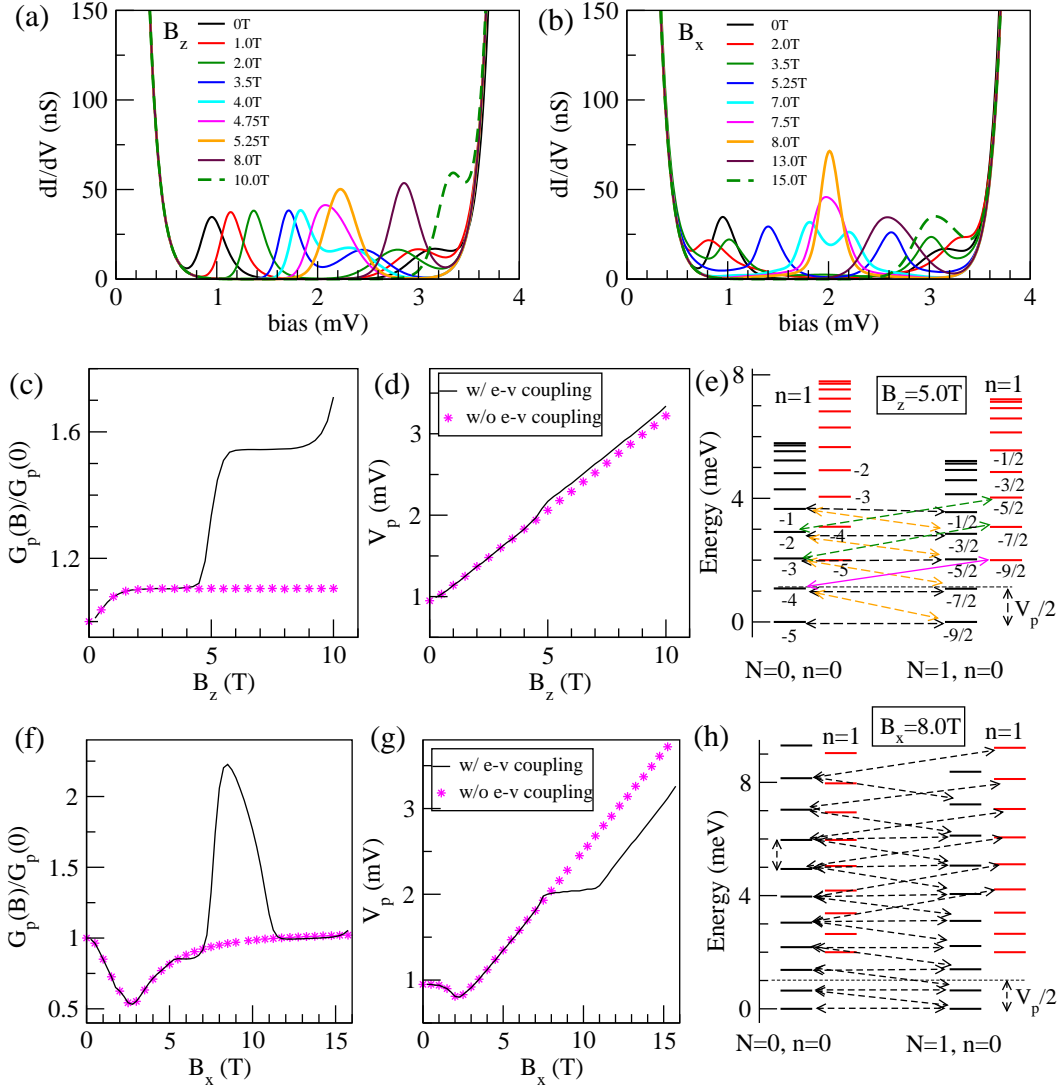

\includegraphics[width=0.85\textwidth]{satellite_peak_0912_nmodes_1.eps}
\includegraphics[width=0.85\textwidth]{satellite_peak_0912_fgh_nmodes_1.eps}
\caption{(Color online) (a) $B_z$-field and (b) $B_x$-field evolution of the satellite conductance peaks between the first and second
main peaks. (c) and (d) Height and position of the leftmost satellite peak vs $B_z$ with and without the electron-vibron coupling.
(e) Dominant tunneling pathways for the leftmost satellite peak at $B_z=5.0$~T, where the solid magenta arrow indicates the transition
determining the peak bias. (f) and (g) Height and position of the leftmost satellite peak vs $B_x$ with and without the electron-vibron
coupling. (h) Dominant tunneling pathways for the leftmost satellite peak at $B_x=8.0$~T, where the vertical arrows indicate
half of the peak bias.}
\label{fig:Bfield2}
\end{figure}


The $B_z$-field evolution of the satellite peaks is shown in Fig.~\ref{fig:Bfield2}(a).
As $B_z$ increases, interestingly, the leftmost satellite peak and the satellite peak on the right side of the second main peak move toward
a higher bias voltage, while the satellite peak on the left side of the second main peak is shifted toward a lower bias voltage. Starting
from the leftmost one, the satellite peaks are referred to as first, second, and third. Around 4.5~T, the first and the second satellite peaks
merge into one peak, and the merged peak moves toward a higher bias voltage. The merged satellite peak disappears above 10.0~T.


Let us focus on the field evolution of the heights and positions of the first and the second satellite peaks. For $B_z > 0.5$~T, the peak bias
$V_p$ for the leftmost satellite peak is dictated by the separation between the two lowest levels $M=-4$ and $M=-5$ (or $m=-7/2$ and $m=-9/2$) 
in the $n=0$ state, which grows linearly with $B_z$, i.e., $V_p(B)/2 = \mathrm{min}\{(9D_0+g\mu_B |B_z|),(8D_1+g\mu_B |B_z|)\}$, as shown in
Fig.~\ref{fig:Bfield2}(d). However, the second satellite peak is governed by a bias voltage where a few low-lying levels in the $n=1$ state
are just about to be populated. The low-lying levels of the $n=1$
state become closer to the first-excited level in the $n=0$ state, as $B_z$ increases. Therefore, with an increase of $B_z$, a smaller bias
voltage can induce a tiny occupation in the low-lying levels of the $n=1$ state, shifting the position of the second satellite peak to the
opposite direction to the first satellite peak. More specifically, for $B_z \lesssim 4.0$~T, the tunneling between the levels in the $n=0$ state
and the levels $n=1$ states is prevented within the first satellite peak bias. However, at $B_z \gtrsim 4.5$~T, several low-lying
levels in the $n=0$ and $n=1$ states are sufficiently close to one another, and so the transitions ($n=0$,$n^{\prime}=1$) are allowed within
the bias window [Fig.~\ref{fig:Bfield2}(e)]. Thus, for $B_z \sim 4.5$~T, the first and second satellite peaks merge, and the transitions
($n=0,n^{\prime}=1$) begin to significantly contribute to the merged satellite peak in addition to the transitions ($n=0,n^{\prime}=0)$. For
higher $B_z$ fields, the contributions of the transitions ($n=0,n^{\prime}=1$) to the merged peak outweigh those of the transitions
($n=0,n^{\prime}=0)$. This explains the abrupt large increase of the height of the merged peak and the sudden small jump in the intercept of the
$V_p$ curve starting from 4.5-5.0~T. Thus, the position and the height of the leftmost satellite peak become largely deviated from the
case of without electron-vibron coupling, as shown in Figs.~\ref{fig:Bfield2}(c) and (d).

We can estimate the $B_z$ value from which the satellite peaks begin to merge. According to the analysis of the transitions ($n=0,n^{\prime}=1$)
similar to that in Sec.V.B.1, at low and intermediate $B_z$ fields, the transitions (0$\rightarrow$1) and (0$\leftarrow$1)
contribute more than the transitions (1$\rightarrow$0) and (1$\leftarrow$0), within the bias window [Fig.~\ref{fig:Bfield1}(c)]. Therefore,
the minimum $B_z$ value where the satellite peaks merge can be determined by the minimum bias window which allows the transition between the
level $M=-4$ in the $n=0$ state and the level $m=-9/2$ in the $n=1$ state, that is, $B_z=[\hbar \omega/2 - 9D_0]/g\mu_B$
[the solid arrow in Fig.~\ref{fig:Bfield2}(e)].
This value is 4.3~T, which agrees with what we find from the actual calculation of the $dI/dV$ vs $V$. The merged satellite peak,
however, disappears when the spacing between the two lowest levels for a given $N$ and $n$ state is comparable to $\hbar \omega$, since in this
case the second main peak appears at the same bias voltage. Even though the first-excited level in the $n=0$ state is degenerate with the
lowest-level in the $n=1$ state at 12.9~T, the merged satellite peak cannot be identified above 10.0~T due to the broadening of the second
main peak.

\subsubsection{$B_x$-field dependence of satellite peaks}

With a $B_x$ field, similarly to the case of $B_z$ field, the first and second satellite peaks are shifted toward the opposite directions, merging
into one, until the merged peak disappears $B_x \gtrsim $16.0~T, as shown in Fig.~\ref{fig:Bfield2}(b). However, the leftmost satellite peak
has distinctive features from the case of $B_z$ field: (1) The peak height forms a large protrusion for 7.5~$ \lesssim B_x \lesssim $~11.0~T 
after which it decreases to the $G_p(0)$ value; (2) The peak voltage remains almost flat for 7.0~$ \lesssim B_x < $~11.0~T; (3) The peak 
disappears at a higher $B_x$ field than in the case of $B_z$ field. Compare Figs.~\ref{fig:Bfield2}(f) and (g) with Figs.~\ref{fig:Bfield2}(c) 
and (d).

We discuss the leftmost satellite peak first for $B_x \lesssim $~11.0~T and then for higher $B_x$ fields. The unique features of the $B_x$ dependence
can be understood by examining how the $B_x$-field evolution of the magnetic levels affects the satellite peaks. For $B_x \leq 2.0$~T, several
low-lying levels are still degenerate [Fig.~\ref{fig:Bfield1x}(f)], and the first satellite peak occurs when a bias voltage is twice as large as
the separation between the two lowest doublets in the $n=0$ state. For such low $B_x$ fields, this separation decreases with increasing $B_x$,
and so do the height and bias voltage of the peak. However, above 3.0~T, the degeneracy of all the levels is lifted, and the peak voltage is much
greater than twice the separation between the two lowest levels in the $n=0$ state. This implies that above 3.0~T, within the bias window,
high-energy levels in the $n=0$ state significantly contribute to the tunneling and the peak height increases with increasing $B_x$. When $B_x$
is increased above 7.5~T, some levels in the $n=0$ and $n=1$ states appear close to one another [Fig.~\ref{fig:Bfield2}(h)], and they can be
accessible within the bias window. The transitions ($n=0,n^{\prime}=1$) are now allowed in the tunneling. Then the first and second satellite peaks
merge and the transitions ($n=0,n^{\prime}=1$) dominantly contribute to the merged peak in addition to ($n=0$,$n^{\prime}=0$). We observe the
sudden large increase in the peak height. The peak height and position are strikingly deviated from those in the case of without
electron-vibron coupling.

However, the trend of the peak height drastically changes, as $B_x$ field increases even further, in contrast to the case of $B_z$. The level
spacing continues to grow with increasing $B_x$. For $B_x > $11.0~T, the separation is so large that the intermediate-energy levels in the $n=0$
and $n=1$ states used to be accessible at lower $B_x$ do not participate in the tunneling anymore within the bias window. Hence, the contributions of
both the transitions ($n=0$,$n^{\prime}=1$) and ($n=0,n^{\prime}=0$) to the peak are highly reduced. Therefore, the peak height drops abruptly, and
the peak position is about twice as large as the spacing between the two lowest levels in a given $N$ and $n$ state. Similarly
to the case of $B_z$, when the spacing between the two lowest levels in the $n=0$ state is comparable to $\hbar \omega$, the merged satellite peak
disappears. With $B_x \neq 0$, the former situation occurs around 18.0~T. Due to the broadening of the second main peak, the satellite peak is not
distinguishable above 16.0~T.

\subsection{Case (ii): Basic features}


\begin{figure}[h]
\includegraphics[width=0.9\textwidth]{nmodes_3_1109.eps}
\includegraphics[width=0.85\textwidth]{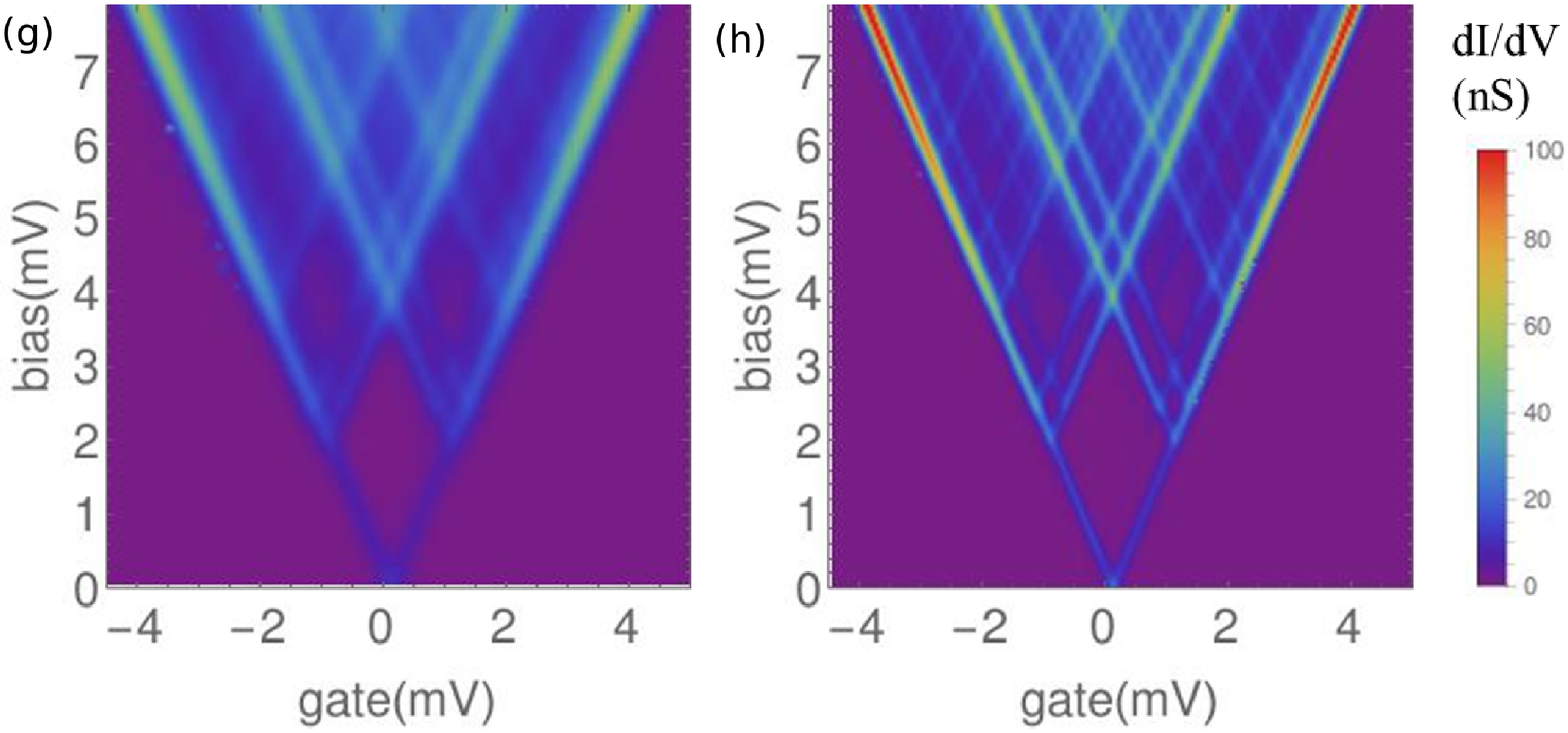}
\caption{(Color Online) Calculated $I-V$ and $dI/dV$ vs $V$ at the charge degeneracy point for the case (ii) at 1.16~K [(a),(b)]
and 0.58~K [(c),(d)]. The arrows in (d) point out the peaks ${\cal M}_2$, ${\cal M}_3$, ${\cal M}_6$, ${\cal M}_7$, ${\cal M}_9$,
and ${\cal M}_{11}$ in Table~\ref{table2}. (e) Magnetic levels in the vibrational ground ($n=0$) and excited states $p_1$, $p_2$, $p_3$,
$q_5$, and $q_6$ for the $N=0$ and $N=1$ states. The excited states are defined in Table~\ref{table1}. Some dominant tunneling
pathways for the second peak at $V/2=2.0$~mV are indicated as the dashed arrows. Not all dominant pathways are drawn for clarity.
(f) Contributions of different transitions $j$ to the second and third peaks. Calculated $dI/dV$ maps as a function of $V$ and $V_g$
for the case (ii) at 1.16~K (g) and 0.58~K (h).}
\label{fig:Tdep_3}
\end{figure}

Figures~\ref{fig:Tdep_3}(a)-(d) exhibit the $I-V$ curve and $dI/dV$ vs $V$ for the case (ii) at the charge degeneracy point for 1.16~K and
0.58~K, respectively. In contrast to the case (i), we find that the current is significantly suppressed at a low bias, and that the $dI/dV$ peak
at zero bias is considerably lower than the peaks arising from vibrational excitations marked by arrows in Figs.~\ref{fig:Tdep_3}(b) and (d).
Compare Fig.~\ref{fig:Tdep_3}(d) with \ref{fig:Tdep_1}(d).
This feature is found at both 1.16~K and 0.58~K. Henceforth, we consider the case at 0.58~K. The peaks from vibrational excitations occur at
$V=$4.0, 5.0, 7.5, 8.0, 9.0, and 10.0~mV for $0 \le V \le 10.0$~mV, which correspond to
$2 \hbar \omega_1$, $2 \hbar \omega_2$, $2 \hbar \omega_3$, $4 \hbar \omega_2$, $2(\hbar \omega_1 + \hbar \omega_2)$, and $4 \hbar \omega_2$,
respectively. In general, peaks from vibrational excitations are found at $V=\sum_{i=1}^3 2 n_i \hbar \omega_i$, where $n_1+n_2+n_3=n > 0$.
All possible vibrational states for $n=0, 1, 2, 3$ are listed in Table~\ref{table1}. Each of the peaks at $V=$4.0, 5.0, 7.5, 8.0, 9.0, and
10.0~mV dominantly originates from transitions between the vibrational ground state and a vibrational excited state, as shown in
Table~\ref{table2}. Among the six peaks, the peak at $V=4.0$~mV has the largest
height. In the bias range of interest, except for the zero-bias peak, four additional main peaks are identified at $V=$6.0, 6.5, 8.5, and
9.5~mV [Fig.~\ref{fig:Tdep_3}(d)], each of which arises dominantly from transitions between a vibrational excited state to another vibrational
excited state, as listed in Table~\ref{table2}. The heights of these peaks are smaller than those of the previous six peaks, because the
vibrational excited states are poorly occupied. 
The stability diagrams shown in Figs.~\ref{fig:Tdep_3}(g) and (h) also support the suppression of the low-bias
current and its robustness with varying $V_g$ and $T$. The diagrams clearly reveal the peaks from the vibrational excitations
parallel to the Coulomb diamond edges in the conduction region. Note that the values of $\lambda_{1,2,3}$ do not differ much from the value
of $\lambda$ for the case (i), and that the ratio of the Franck-Condon factor for the peak at 4.0~mV to the factor at zero bias is the same
for both cases, such as ${\cal F}_{\mathbf{0},p_3}/{\cal F}_{\mathbf{0},\mathbf{0}}={\cal F}_{0,1}/{\cal F}_{0,0}=\lambda^2=\lambda_1^2$.
(Several values of the Franck-Condon factor for the case (ii) are listed in Table~\ref{table4} in the Appendix.) Nonetheless, the case (ii)
produces an effect similar to what was shown for a single mode with stronger electron-vibron coupling, referred to as the Franck-Condon blockade
effect \cite{KOCH05,KOCH06}.


\begin{table}
\begin{center}
\caption{List of all vibrational states $(n_1,n_2,n_3)$ and their energies $E_{(n_1,n_2,n_3)}$ (meV) for $n=0, 1, 2, 3$, where
$E_{(n_1,n_2,n_3)}=\sum_{i=1}^3 n_i \hbar \omega_i$.}
\label{table1}
\begin{ruledtabular}
\begin{tabular}{c|c|c|c|c|c|c}
label        & $n_1$ & $n_2$ & $n_3$ & $n$ & $E_{(n_1,n_2,n_3)}$ & $2 E_{(n_1,n_2,n_3)}$ \\ \hline
$\mathbf{0}$ &  0    &   0   &   0   &  0  &  0    &  0 \\
$p_1$        &  0    &   0   &   1   &  1  & 3.75  &  7.5 \\
$p_2$        &  0    &   1   &   0   &  1  & 2.5   &  5.0 \\
$p_3$        &  1    &   0   &   0   &  1  & 2.0   &  4.0 \\
$q_1$        &  0    &   0   &   2   &  2  & 7.5   &  15.0 \\
$q_2$        &  0    &   1   &   1   &  2  & 6.25  &  12.5 \\
$q_3$        &  0    &   2   &   0   &  2  & 5.0   &  10.0 \\
$q_4$        &  1    &   0   &   1   &  2  & 5.75  &  11.5 \\
$q_5$        &  1    &   1   &   0   &  2  & 4.5   &  9.0 \\
$q_6$        &  2    &   0   &   0   &  2  & 4.0   &  8.0 \\
$r_1$        &  0    &   0   &   3   &  3  & 11.25 & 22.5 \\
$r_2$        &  0    &   1   &   2   &  3  & 10.0  & 20.0 \\
$r_3$        &  0    &   2   &   1   &  3  & 8.75  & 17.5 \\
$r_4$        &  0    &   3   &   0   &  3  & 7.5   & 15.0 \\
$r_5$        &  1    &   0   &   2   &  3  & 9.5   & 19.0 \\
$r_6$        &  1    &   1   &   1   &  3  & 8.25  & 16.5 \\
$r_7$        &  1    &   2   &   0   &  3  & 7.0   & 14.0 \\
$r_8$        &  2    &   0   &   1   &  3  & 7.75  & 15.5 \\
$r_9$        &  2    &   1   &   0   &  3  & 6.5   & 13.0 \\
$r_{10}$     &  3    &   0   &   0   &  3  & 6.0   & 12.0 \\
\end{tabular}
\end{ruledtabular}
\end{center}
\end{table}

\begin{center}
\begin{table}
\caption{Eleven identified main $dI/dV$ peaks shown in Fig.~\ref{fig:Tdep_3}(d) with the peak voltages $V_p$ (in meV) and the dominant
transitions. The peak ${\cal M}_{11}$ at 10.0~mV arises from equally dominant two transitions $(p_2,q_1)$ and ($\mathbf{0}$,$q_3$).
Here ($\mathbf{0},p_3$) represents all allowed transitions such as
$\{ |\psi^{N=0}_{M} \rangle \otimes | \mathbf{0} \rangle \leftrightarrow |\psi^{N=1}_{m} \rangle \otimes | p_3 \rangle \}$
and $\{ |\psi^{N=0}_{M} \rangle \otimes | p_3 \rangle \leftrightarrow |\psi^{N=1}_{m} \rangle \otimes | \mathbf{0} \rangle \}$.
Refer to Table~\ref{table1} for the definitions of the vibrational states in the dominant transitions.}
\label{table2}
\begin{ruledtabular}
\begin{tabular}{cccccccccccc}
Label & ${\cal M}_1$ & ${\cal M}_2$ & ${\cal M}_3$ & ${\cal M}_4^{\star}$ & ${\cal M}_5^{\star}$ &
${\cal M}_6$ & ${\cal M}_7$ & ${\cal M}_8^{\star}$ & ${\cal M}_9$ & ${\cal M}_{10}^{\star}$ & ${\cal M}_{11}$ \\ \hline
$V_p$ & 0 & 4.0 & 5.0 & 6.0 & 6.5 & 7.5 & 8.0 & 8.5 & 9.0 & 9.5 & 10.0 \\
Tran. & ($\mathbf{0}$,$\mathbf{0}$) & ($\mathbf{0}$,$p_3$) & ($\mathbf{0}$,$p_2$) & ($p_3$,$q_3$) & ($p_2$,$q_4$) &
($\mathbf{0}$,$p_1$) & ($\mathbf{0}$,$q_6$) & ($p_3$,$q_2$) & ($\mathbf{0}$,$q_5$) & ($q_6$,$r_3$) & ($\mathbf{0}$,$q_3$) \\
\end{tabular}
\end{ruledtabular}
\end{table}
\end{center}

To analyze the $dI/dV$ peak height, we separate contributions of different transitions $j$ to the first, second, and third main peaks
(${\cal M}_1$, ${\cal M}_2$, ${\cal M}_3$) at 0, 4.0, and 5.0~mV. As shown in Fig.~\ref{fig:Tdep_3}(f), at $V=4.0$~mV, the height of the
peak ${\cal M}_2$ arising solely from transitions ($\mathbf{0}$,$\mathbf{0}$), ($\mathbf{0},p_3$), and ($p_3$,$p_3$), is about 9.5~nS,
and this height is smaller than the height of the zero-bias peak ${\cal M}_1$ ($\sim$22~nS). The fact that the former height is smaller than the
latter height, is similar to the case (i). However, interestingly, transitions ($p_1$,$p_3$), ($p_2$,$p_3$), ($p_1$,$p_2$), and ($p_1$,$q_5$)
considerably contribute to the peak ${\cal M}_2$ with additional 22 nS, and so the total height of the peak ${\cal M}_2$ becomes larger
than the height of the peak ${\cal M}_1$. This strikingly differs from the case (i) where the transitions ($n=1$,$n^{\prime}=1$) provide
only a tiny increase of the height of the second peak (Fig.~\ref{fig:Tdep_1}(f) in Sec.V.A).
The key difference between the cases (i) and (ii) is that the latter has two additional modes whose energies are close
to that of the lowest-energy mode. At $V/2=\hbar \omega_1$, the lowest levels $M=\pm 5$ or $m=\pm 9/2$ in the $p_3$ state 
are significantly occupied. Hence, for $(\hbar \omega_2 - \hbar \omega_1) < \hbar \omega_1$ and $(\hbar \omega_3 - \hbar \omega_1) < \hbar \omega_1$, 
the transitions ($p_1$,$p_3$) and ($p_2$,$p_3$) are also allowed [Fig.~\ref{fig:Tdep_3}(e)]. Accordingly, the levels $M=\pm 5$
or $m=\pm 9/2$ in the $p_2$ and $p_1$ states are somewhat occupied, and so the transitions ($p_1$,$p_2$) and ($p_1$,$q_5$) are possible 
within the bias window. A similar analysis can be carried out for the
third peak height. In this case, at $V=5.0$~mV, transitions ($\mathbf{0},p_2$) play a major role in the peak height, while transitions
($p_1$,$p_2$), ($p_2$,$p_3$), ($p_2$,$q_6$), and ($p_1$,$q_5$) provide considerable contributions to the height [Fig.~\ref{fig:Tdep_3}(f)].
The overall height of the third peak turns out to be greater than the height of the zero-bias peak, although it is smaller than the second peak
height. Similarly to the case (i), a satellite peak occurs at 1.0 mV and a flat shoulder appears on the left side of the second main peak
[Fig.~\ref{fig:Tdep_3}(d)], which is attributed to the magnetic anisotropy. The first (leftmost)
satellite peak can be explained similarly to the case (i).

The Franck-Condon blockade effect has recently been observed in single-molecule transistors made of individual Fe$_4$ molecules \cite{BURZ14},
where the experimental data were fitted to vibrational excitations from a single normal mode of a non-magnetic molecule. The experimental
values of $\lambda$ and $\hbar \omega$ were 2.0$\pm$0.2 and 2.3-2.6~meV, respectively \cite{BURZ14}, and they are in reasonable agreement
with the corresponding DFT-calculated values. With the experimental level broadening
$\Gamma \sim 1.0$~meV, the vibrational excitations may not be individually identified, and the calculated peaks at 4.0 mV and 5.0 mV could be
viewed as a single peak in the experimental data.

\subsection{Case (ii): Magnetic field dependence}

\begin{figure}[h]
\includegraphics[width=0.85\textwidth]{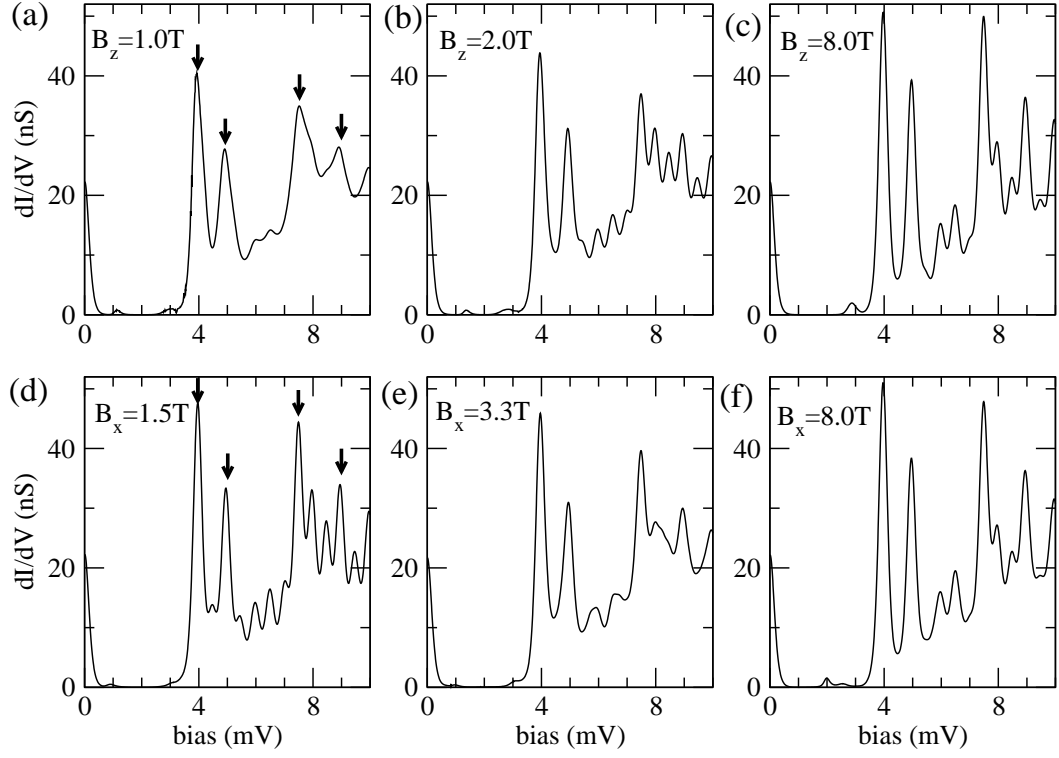}
\caption{(a)-(c) $B_z$-field and (d)-(f) $B_x$-field dependence of $dI/dV$ vs $V$ in the case (ii) for the charge degeneracy point at 0.58~K.
The arrows in (a) and (d) correspond to the peaks ${\cal M}_2$, ${\cal M}_3$, ${\cal M}_6$, and ${\cal M}_9$ in Table~\ref{table2}.}
\label{fig:BBB}
\end{figure}

The heights of the main peaks and the heights and positions of the satellite peaks show strong $B$-field dependencies (Fig.~\ref{fig:BBB}),
while the positions of the main peaks relative to the charge degeneracy point do not change with $B$ field, which is similar to the case (i).
The main peaks from the tunneling between the levels in the $n=0$ state and the low-energy vibrational excited state, such as ${\cal M}_2$,
${\cal M}_3$, ${\cal M}_6$, and ${\cal M}_9$ (marked by the arrows in Fig.~\ref{fig:BBB}), have still a larger height than the zero-bias
peak, independently of the orientation and magnitude of $B$ field. Some main peaks involved with either high-energy vibrational excited states
or close to the other main peaks, are smeared out at some $B$ fields. For example, the three peaks ${\cal M}_7$, ${\cal M}_8^{\star}$, and
${\cal M}_{10}^{\star}$ used to appear at $V=$8.0, 8.5, and 9.5~mV in the absence of $B$ field, respectively, are not found at $B_z=1.0$~T
[Fig.~\ref{fig:BBB}(a)]. In addition, the peaks ${\cal M}_8^{\star}$ and ${\cal M}_{10}^{\star}$ occurring at 8.5 and 9.5~mV for $B=0$,
are not apparent for $B_x=3.3$~T, as shown in Fig.~\ref{fig:BBB}(e). In this section, we focus on the first, second, and third main
peaks (${\cal M}_1$, ${\cal M}_2$, ${\cal M}_3$)and the satellite peaks between the first and second main peaks, in the presence of 
$B_z$ or $B_x$ field at 0.58~K for the charge degeneracy point. The height of the third main peak reveals a $B$-field dependence 
qualitatively different from that in the case (i). Note that the former peak dominantly arises from the tunneling between the levels 
in the $n=0$ state and in one of the $n=1$ states ($p_2$ state), while the latter peak mainly originates from the tunneling between the 
levels in the $n=0$ and $n=2$ states. Since some conductance features in the case (ii) are similar to those in the case (i), we underscore 
results distinctive from the case (i).

\subsubsection{Main peaks}

\begin{figure}[h]
\includegraphics[width=0.95\textwidth]{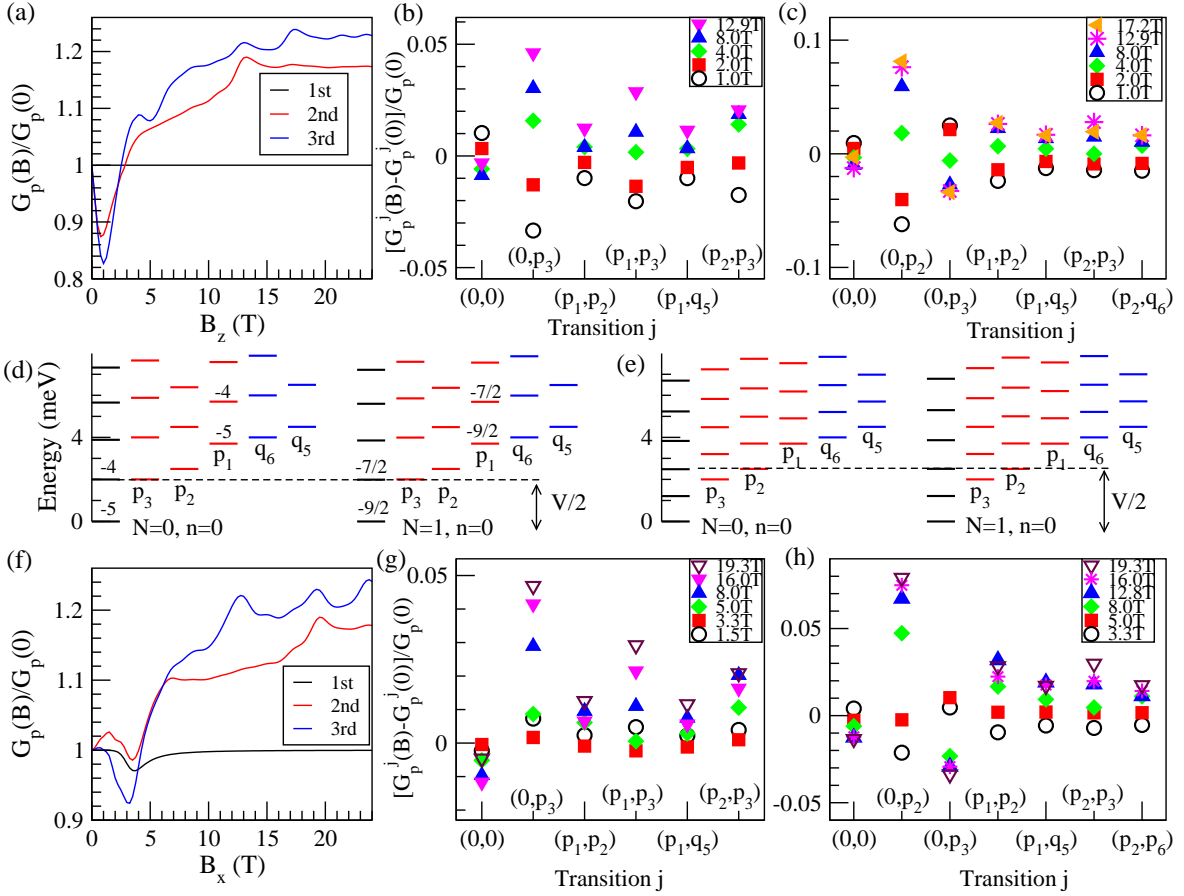}
\caption{(Color online) (a) Ratio $G_p(B)/G_p(0)$ for the first (black), second (red), and third (blue) main peaks vs $B_z$ field for the
case (ii) at 0.58~K.
Normalized changes of contributions of different transitions $j$, $[G_p^j(B_z)-G_p^j(0)]/G_p(0)$, for the second (b) and the third peaks
(c) at different $B_z$ fields. Magnetic levels of the $n=0$, $p_1$, $p_2$, $p_3$, $q_5$, and $q_6$ states for the $N=0$ and $N=1$ states at
$B_z=12.9$~T (d) and at $B_x=12.8$~T (e). Only the low-energy levels are shown to emphasize the low-energy region. The vertical arrow in
(d) represents half of the bias voltage for the second peak with $B_z$ field, while the vertical arrow in (e) indicates half of the 
bias voltage for
the third peak with $B_x$ field. (f) Ratio $G_p(B)/G_p(0)$ for the first, second, and third main peaks vs $B_x$ field for the case (ii) at
0.58~K. Changes of contributions of transitions $j$, $[G_p^j(B_z)-G_p^j(0)]/G_p(0)$, at various $B_x$ fields for the second (g) and 
the third peaks (h).}
\label{fig:Bfield3}
\end{figure}


As $B_z$ field increases, the heights of the second and third main peaks sharply decrease near 1.0~T, and then they rapidly rise well above the
$G_p(B=0)$ values [Fig.~\ref{fig:Bfield3}(a)]. This is in contrast to the case (i), where the heights remain saturated to much lower
values than the $G_p(0)$ values until about 12.0~T. Compare Fig.~\ref{fig:Bfield3}(a) with Fig.~\ref{fig:Bfield1}(a). To understand this
difference, we examine $[G_p^j(B_z)-G_p^j(0)]/G_p(0)$ for different transitions $j=(n,n^{\prime})$. At 1.0~T, similarly to the case (i),
the abrupt drops of the heights of the peaks are due to the lift of the level degeneracy, which brings a large decrease of the dominant
transitions ($\mathbf{0},p_3$) at $V=4.0$~mV and a large decrease of ($\mathbf{0},p_2$) for $V=5.0$~mV, as shown in
Fig.~\ref{fig:Bfield3}(b) and (c). However, as $B_z$ field increases, the transitions ($\mathbf{0},p_3$) [($\mathbf{0},p_2$)] and other
transitions begin to contribute more to the second (third) peak than at zero $B$ field, since new tunneling pathways are available from the
three vibrational modes, compared to the case (i). More specifically, within $V=4.0$~mV, the transitions ($\mathbf{0},p_3$) and ($p_2$,$p_3$) participate in the tunneling more at 4.0~T than at zero $B$, while the transitions ($\mathbf{0},p_3$) and ($p_1$,$p_3$) involve more at
8.0~T than at 4.0~T. Within $V=5.0$~mV, the transitions ($\mathbf{0},p_2$) contribute to the third peak more at 4.0~T than at zero $B$,
while the transitions ($\mathbf{0},p_2$), ($p_1$,$p_2$), and ($p_2$,$p_3$) participate in the peak more at 8.0~T than at 4.0~T.

The small bumps in the heights of the second and third peaks at $B_z=$12.9~T appear due to the same reason as in the case (i) (Sec.V.B.1).
At this $B_z$ field, the spacing between the lowest and the first-excited levels for a given $N$ and $n$ state is comparable to
$\hbar \omega_1$, such that the first-excited level in the $n=0$ state is degenerate with the lowest level in the $p_3$ state [Fig.~\ref{fig:Bfield3}(d)]. For $V=2 \hbar \omega_1$, this noticeably increases the occupation of these two levels and the lowest level
in the $n=0$ state, giving rise to an additional boost of the contributions of ($\mathbf{0},p_3$) and ($p_1$,$p_3$) to the second peak
compared to lower $B_z$ fields. For $V=2 \hbar \omega_2$, there is an increase of the occupation of the four levels in each charge state
within the bias window, leading to an increase of the contributions of ($\mathbf{0},p_2$) and ($p_2$,$p_3$) to the third peak. Compare the
peak height differences at 12.9~T with those at 8.0~T in Figs.~\ref{fig:Bfield3}(b) and (c). Another bump in the height of
the third peak occurs at 17.2~T, where the spacing between the lowest and the first-excited levels for a given $N$ and $n$ state is
now comparable to $\hbar \omega_2$, i.e., $(\hbar \omega_2/2 - 9D_0)/g\mu_B=17.2$~T. In this case, the first-excited level in the $n=0$
state is degenerate with the lowest level in the $p_2$ state, which results in a slight increase of the transitions
($\mathbf{0},\mathbf{0}$) and ($p_1$,$q_5$), within $V = 2 \hbar \omega_2$, compared to lower $B_z$ fields (not shown). This produces
a small bump in the third peak height at 17.2~K.


With a $B_x$ field, the third peak height shows an interesting feature, although the field dependence of the heights of the first and second
main peaks is similar to that for the case (i). The height of the third main peak drops until 3.3~T, and as $B_x$ increases, it goes up with three
apparent bumps at 12.8, 19.3, and 23.8~T, as shown in Fig.~\ref{fig:Bfield3}(f). At the first bump, the spacing between the second-excited level
and the lowest level for a given $N$ and $n$ state is comparable to $\hbar \omega_2$, and so the second-excited level in the $n=0$ state is degenerate
with the lowest level in the $p_2$ state [Fig.~\ref{fig:Bfield3}(e)]. For $V= 2 \hbar \omega_2$, this provides a substantial increase of the occupation of the levels within the bias window, leading to an increase of the transitions ($\mathbf{0},p_2$), ($p_1$,$p_2$) and ($p_2$,$p_3$), as indicated in Fig.~\ref{fig:Bfield3}(h). At the second bump, similarly to the case (i), the first-excited level in the $n=0$ state has the same energy as the lowest level in the $p_3$ state, giving rise to a slight increase of contributions of the transitions ($p_2$,$p_3$) to the third peak, compared to lower
$B_x$ fields. At this $B_x$ field, a bump also appears in the height of the second peak [Fig.~\ref{fig:Bfield3}(g)], similarly to the case (i) (Sec.V.B.2). At the third bump, the first-excited level in the $n=0$ state
is degenerate with the lowest level in the $p_2$ state. A slight increase of transitions ($\mathbf{0},\mathbf{0}$) brings the small
bump in the third peak.

\subsubsection{Satellite peaks}

\begin{figure}[h]
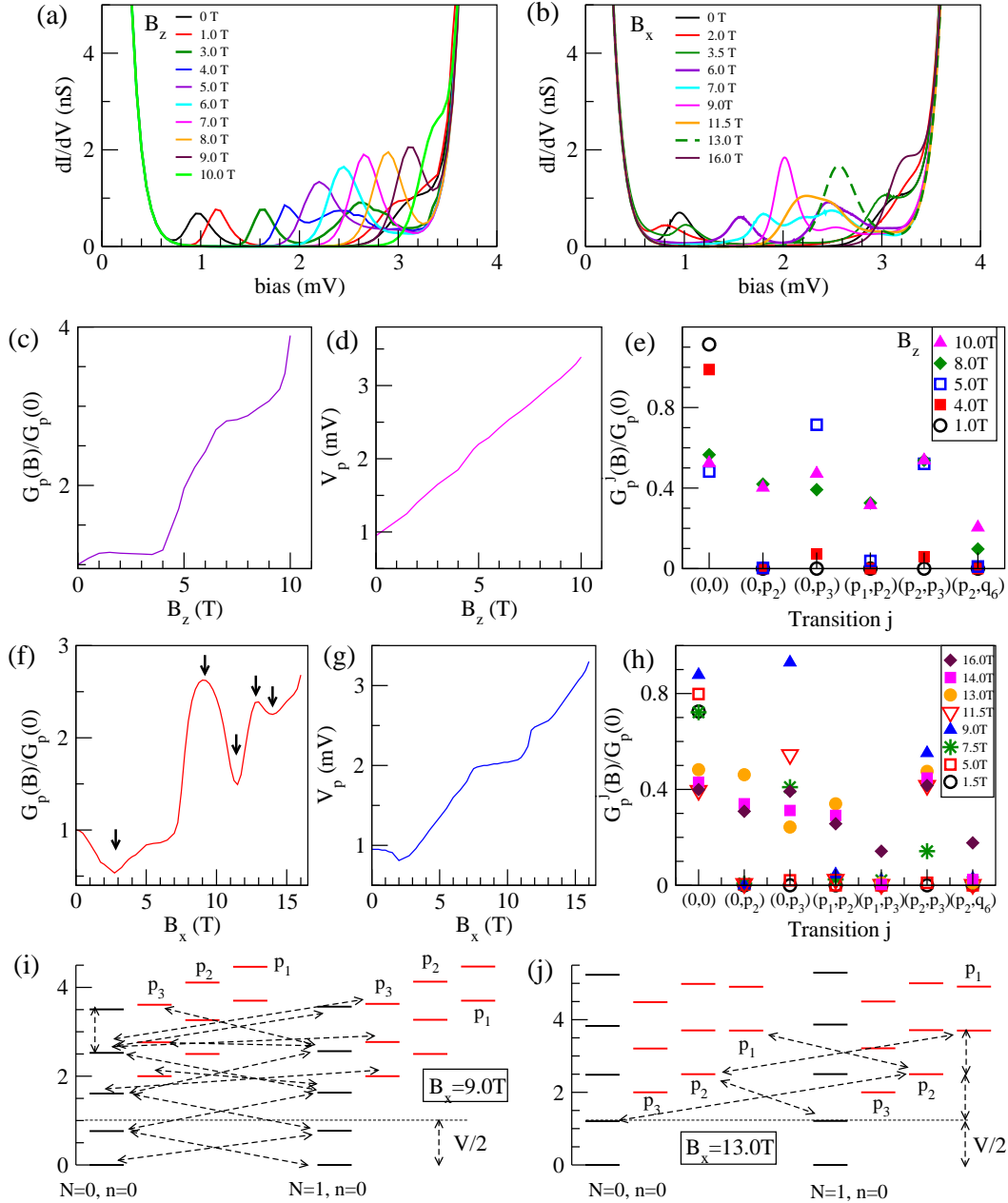

\includegraphics[width=0.85\textwidth]{satellite_peak_0911.eps}
\includegraphics[width=0.85\textwidth]{satellite_peak_0910_fgh.eps}
\caption{(Color online) (a) $B_z$-field and (b) $B_x$-field evolution of the satellite peaks for the case (ii) at 0.58~K. The height (c) and
position (d) of the leftmost satellite peak vs $B_z$ field. (e) $G_p^{j}(B)/G_p(0)$ vs transitions $j$ for the leftmost satellite peak at various
$B_z$ fields. The height (f) and position (g) of the leftmost satellite peak vs $B_x$ field. In (f), the arrows indicate the local minima and
maxima in the height of the leftmost satellite peak. (h) $G_p^{j}(B)/G_p(0)$ vs transitions $j$ for the leftmost satellite peak at various 
$B_x$ fields. Dominant tunneling pathways for the leftmost satellite peak at $B_x$=9.0~T (i) and 13.0~T (j), where the vertical arrows 
indicate half of the peak bias. Only low-energy magnetic levels are shown to emphasize the transitions in the low-energy region.}
\label{fig:Bfield4}
\end{figure}

We first discuss the case of $B_z$ field. Figure~\ref{fig:Bfield4}(a) shows how the satellite peaks between the first and second main peaks
evolve with $B_z$ field. Compare Figs.~\ref{fig:Bfield4}(a),(c),(d) with Figs.~\ref{fig:Bfield2}(a),(c),(d). Similarly to the case (i), near
4.5~T, the leftmost satellite peak merges with the second satellite peak, and the merged satellite peak has a large height which is strongly
deviated from the case of without electron-vibron coupling. We find that at 5.0~T, the transitions ($\mathbf{0}$,$p_3$) and ($p_2$,$p_3$)
contribute to the first satellite peak as much as the transitions ($\mathbf{0}$,$\mathbf{0}$) [Fig.~\ref{fig:Bfield4}(e)]. The energy difference between the lowest levels in the $p_3$ and $p_2$ states is only 0.5~meV, and the low-lying
levels in the $p_3$ state are occupied from the transitions ($\mathbf{0}$,$p_3$). Thus, the transitions ($p_2$,$p_3$) can participate in the
tunneling for a bias window of 2.2~mV at 5.0~T. As $B_z$ field further increases, more diverse types of transitions contribute to the merged
satellite peak. Interestingly, the contributions of the transitions ($p_1$,$p_3$) to the merged satellite peak are negligible, because the 
energy difference between
the lowest levels in the $p_3$ and $p_1$ states exceeds a half of the peak bias voltage. The transitions ($p_1$,$q_5$) do not contribute to the
satellite peak despite the small energy difference between the lowest levels in the $p_1$ and $q_5$ state because the levels in the $p_1$ state
are not occupied. The merged peak height increases to a much higher value than the case (i), although the peak position is the same as
that for the case (i). The merged satellite peak eventually disappears above 10.0~T at 0.58~K, attributed to the same reason as in the case
(i) (Sec.V.B.3).


With a $B_x$ field, below 7.0~T, the evolution of the satellite peaks [Fig.~\ref{fig:Bfield4}(b)] is similar to the case (i) (Sec.V.B.4).
As $B_x$ field increases, the leftmost satellite peak is shifted toward a higher bias, while the second satellite peak moves toward a lower bias. Interestingly, around 7.0~T, three satellite peaks appear instead of two, while around 9.0~T, the
first two satellite peaks become merged but the third peak still survives [Fig.~\ref{fig:Bfield4}(b)]. Then at 13.0~T, the survived two satellite
peaks are completely merged. The merged peak is shifted toward a higher bias at higher $B$ fields. Compare Fig.~\ref{fig:Bfield4}(b) with
Fig.~\ref{fig:Bfield2}(b) at 7.0~T and 9.0~T. Comparing with the case (i), the leftmost peak height does not drop to the $G_p(0)$ value above
11.5~T. Instead it resumes to grow and reaches to a local maximum at 13.0~T. Then the height undergoes a slight decrease with another upturn until
the merged peak disappears above 16.0~T [Fig.~\ref{fig:Bfield4}(f)]. At the $B_x$ fields where the height of the leftmost satellite peak reaches
to local maxima, the peak position remains almost flat [Fig.~\ref{fig:Bfield4}(g)].

Above $B_x=$7.0~T, the vibrational excited states play an important role even in the satellite peaks. At 7.5~T, the transitions
($\mathbf{0}$,$p_3$) contribute substantially to the leftmost satellite peak, while at 9.0~T (at the maximum peak height),
there is a great increase of the transitions ($\mathbf{0}$,$p_3$) and ($p_2$,$p_3$) compared to lower $B_x$ fields 
[Fig.~\ref{fig:Bfield4}(h)]. The peak bias
at 9.0~T is still higher than twice the energy difference between the lowest levels in the $n=0$ state. The level spacing is larger
for higher levels. Some dominant transition pathways within ($\mathbf{0}$,$p_3$) are shown in Fig.~\ref{fig:Bfield4}(i).
The transitions ($\mathbf{0}$,$\mathbf{0}$) allow the intermediate-energy levels of the $n=0$ state to be occupied, and the transitions
($\mathbf{0}$,$p_3$) occur between these levels and the low-lying levels in the $p_3$ state. Now the occupation in the $p_3$ state
induces the transitions ($p_2$,$p_3$). As $B_x$ increases to 11.5~T, the transitions ($\mathbf{0}$,$p_3$) and ($p_2$,$p_3$) greatly
decrease, which gives rise to a drop in the peak height [Fig.~\ref{fig:Bfield4}(f),(h)]. As discussed in the case (i), above 11.5~T,
the peak bias is determined by the spacing between the two lowest levels in the $n=0$ state. At 13.0~T (at the local maximum height), 
the spacing between the lowest and the first-excited level in the $n=0$ state is comparable to the energy difference between the 
first-excited level in the $n=0$ state and the lowest-level in the $p_2$ state [vertical arrows in Fig.~\ref{fig:Bfield4}(j)].
Thus, at this $B$ field,
the transitions ($\mathbf{0}$,$p_2$) increase and they contribute to the satellite peak height. Moreover, the level spacing in the $n=0$
state is comparable to the energy difference between the lowest levels in the $p_2$ and $p_1$ state. Since the lowest level in the $p_2$ state
is occupied, the transitions ($p_1$,$p_2$) also contribute to the peak height at this $B$ field. Dominant transition pathways among
($\mathbf{0}$,$p_2$) and ($p_1$,$p_2$) are shown in Fig.~\ref{fig:Bfield4}(j). With a higher $B_x$ field, other transitions requiring
higher energies participate in the tunneling. The merged satellite peak disappears above 16.0~T.

\section{Conclusion}

We have shown that magnetic anisotropy provides new features concerning electron-vibron coupling in electron transport through single
anisotropic molecules such as the SMM Fe$_4$. The heights of the vibrational
conductance peaks show an unusual $B$-field dependence at low temperatures. When the current flows via the vibrational excited states of
the Fe$_4$, the magnetic levels in the vibrational ground and excited states participate in the tunneling. The separation between the
magnetic levels strongly depends on the direction and magnitude of applied $B$-field, and so the occupation of the levels and transition
rates between them are accordingly modified with $B$ field.
As a result, the vibrational conductance peaks are highly influenced by the direction as well as magnitude of the applied $B$
field. Interestingly, when the two lowest levels in the $n=0$ state are separated by about the vibrational energies at high $B$ fields,
a sudden large jump in the peak height is expected. Moreover, the magnetic anisotropy introduces satellite conductance peaks whose position
and height are varied with the direction and magnitude of the applied $B$ field. At zero $B$ field, the low-bias satellite peak originates
from the current via the magnetic levels in the vibrational ground states only, while at intermediate $B$ fields, the levels in the
vibrational first-excited state start to contribute to the satellite peak. Another interesting point is the effect of multiple strong
electron-vibron coupled modes whose energies
are close to one another. For such multiple modes the vibrational conductance peaks are greatly enhanced compared to a single mode with
the similar electron-vibron coupling. Our findings may be extended to studies of spin-vibron coupling effects, higher-order tunneling
processes, and many-spin model Hamiltonian in transport via individual anisotropic molecules.
For comparison with experiment, caution has to be exercised in two aspects such as level broadening and excited spin multiplets.
An experimental level broadening may correspond to a sum of the broadenings of several magnetic levels rather than the broadening of
one level. In the case of without electron-vibron coupling, numerical renormalization group studies show that main transport properties
are not affected by consideration of a larger level broadening \cite{MISI14}. Very high external $B$ fields could induce some overlap 
with the low-lying levels of the excited spin multiplet(s) in each charge state.

\begin{acknowledgments}
A.M., Y.Y., M.W., and K.P. were supported by the U. S. National Science Foundation DMR-1206354, SDSC
Trestles under DMR060009N, and Advanced Research Computing at Virginia Tech. E.B. and H.S.J.v.d.Z were supported by
the EU FP7 program through project 618082 ACMOL and an advanced ERC grant (Mols@Mols).
The authors are grateful to J.S. Seldenthuis, J.M. Thijssen, and M. Misiorny for discussions to solve the master equation,
and to A. Cornia for discussion on the vibrational properties of the Fe$_4$.
\end{acknowledgments}

\clearpage

\appendix*

\section{Franck-Condon factors for several transitions in the cases (i) and (ii).}

The Franck-Condon factors for several transitions are computing for the cases (i) and (ii), by applying the recursion relations
\cite{RUHO94,RUHO00} to the overlap matrices (Sec.IV.B).

\begin{center}
\begin{table}[h]
\caption{Franck-Condon factors for several transitions $(n,n^{\prime})$ for the case (i).}
\label{table3}
\begin{ruledtabular}
\begin{tabular}{c|c}
$(n,n^{\prime})$ & ${\cal F}_{n,n^{\prime}}$ \\ \hline
$(0,0)$ & 0.199 \\
$(0,1)$ & 0.321 \\
$(0,2)$ & 0.259 \\
$(0,3)$ & 0.139 \\
$(1,1)$ & 0.075 \\
$(1,2)$ & 0.024 \\
$(1,3)$ & 0.166 \\
$(2,2)$ & 0.171 \\
$(2,3)$ & 0.031 \\
$(3,3)$ & 0.081
\end{tabular}
\end{ruledtabular}
\end{table}
\end{center}

\begin{center}
\begin{table}[h]
\caption{Franck-Condon factors for several transitions $(n,n^{\prime})$ for the case (ii).}
\label{table4}
\begin{ruledtabular}
\begin{tabular}{c|c}
$(n,n^{\prime})$ & ${\cal F}_{n,n^{\prime}}$ \\ \hline
(${\mathbf 0}$,${\mathbf 0}$) & 0.00403 \\
(${\mathbf 0}$,$p_1$) & 0.00860 \\
(${\mathbf 0}$,$p_2$) & 0.00713 \\
(${\mathbf 0}$,$p_3$) & 0.00650 \\
($p_1$,$p_1$) & 0.00516 \\
($p_1$,$p_2$) & 0.0152 \\
($p_1$,$p_3$) & 0.0139 \\
($p_1$,$q_5$) & 0.0245 \\
($p_1$,$q_6$) & 0.0112 \\
($p_2$,$p_2$) & 0.00238 \\
($p_2$,$p_3$) & 0.0115 \\
($p_2$,$q_5$) & 0.00385 \\
($p_2$,$q_6$) & 0.00928 \\
($p_3$,$p_3$) & 0.00151 \\
($p_3$,$q_5$) & 0.00268 \\
($p_3$,$q_6$) & 0.000487
\end{tabular}
\end{ruledtabular}
\end{table}
\end{center}

\end{document}